\documentclass[preprint]{vgtc}               

\graphicspath{{figures/}{pictures/}{images/}{./}} 

\usepackage{microtype}                 
\usepackage{times}                     

\usepackage{tabu}                      
\usepackage{booktabs}                  
\usepackage{lipsum}                    
\usepackage{mwe}                       

\usepackage[svgnames]{xcolor}
\usepackage[linesnumbered,ruled,vlined,commentsnumbered]{algorithm2e}
\usepackage{amsfonts}
\usepackage{amsmath}
\usepackage{tikz}
\usepackage{siunitx}
\usepackage{adjustbox}

\usepackage{algpseudocode}
\usepackage{subcaption}
\usepackage{pdfpages}

\definecolor{lime}{HTML}{A6CE39}
\DeclareRobustCommand{\orcidicon}{%
    \begin{tikzpicture}
    \draw[lime, fill=lime] (0,0) 
    circle [radius=0.16] 
    node[white] {{\fontfamily{qag}\selectfont \tiny \textbf{ID}}};
    \draw[white, fill=white] (-0.065,0.1) 
    circle [radius=0.007];
    \end{tikzpicture}
    \hspace{-2mm}
}
\newcommand{\orcid}[1]{\href{https://orcid.org/#1}{\orcidicon}}

\usepackage{mathptmx}                  

\onlineid{1033}

\vgtccategory{Research}

\vgtcinsertpkg

\title{Topological Simplifcation of Jacobi Sets for Piecewise-Linear Bivariate\\ 2D Scalar Fields with Adjustment of the Underlying Data}

\author{Felix Raith\orcid{0000-0002-3505-6130}\thanks{e-mail: raith@informatik.uni-leipzig.de}\\ %
        \scriptsize Leipzig University %
\and Gerik Scheuermann\orcid{0000-0001-5200-8870}\thanks{e-mail: scheuer@informatik.uni-leipzig.de}\\ %
     \scriptsize Leipzig University \& ScaDS.AI %
\and Christian Heine\orcid{0000-0001-7067-5650}\thanks{e-mail: heine@informatik.uni-leipzig.de}\\ %
     \scriptsize Leipzig University}

\authorfooter{
    \item
        Felix Raith is with Leipzig University.
        E-mail: raith@informatik.uni-leipzig.de.
    \item
        Gerik Scheuermann is with Leipzig University
        E-mail: scheuer@informatik.uni-leipzig.de.

    \item Christian Heine is with Leipzig University.
        E-mail: heine@informatik.uni-leipzig.de.
}

\teaser{
    \centering
    \includegraphics[height=0.447\linewidth, alt={Nice Example to jacobi sets simplification before and after this algorithm.}]{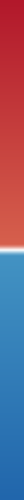}
    \begin{subfigure}[b]{0.9\linewidth}
    \resizebox{!}{0.25\linewidth}{%
    \begin{tikzpicture}
        \node[anchor=south west,inner sep=0] at (0,0)    {\includegraphics[height=0.5\linewidth, alt={Nice Example to jacobi sets simplification before and after this algorithm.}]{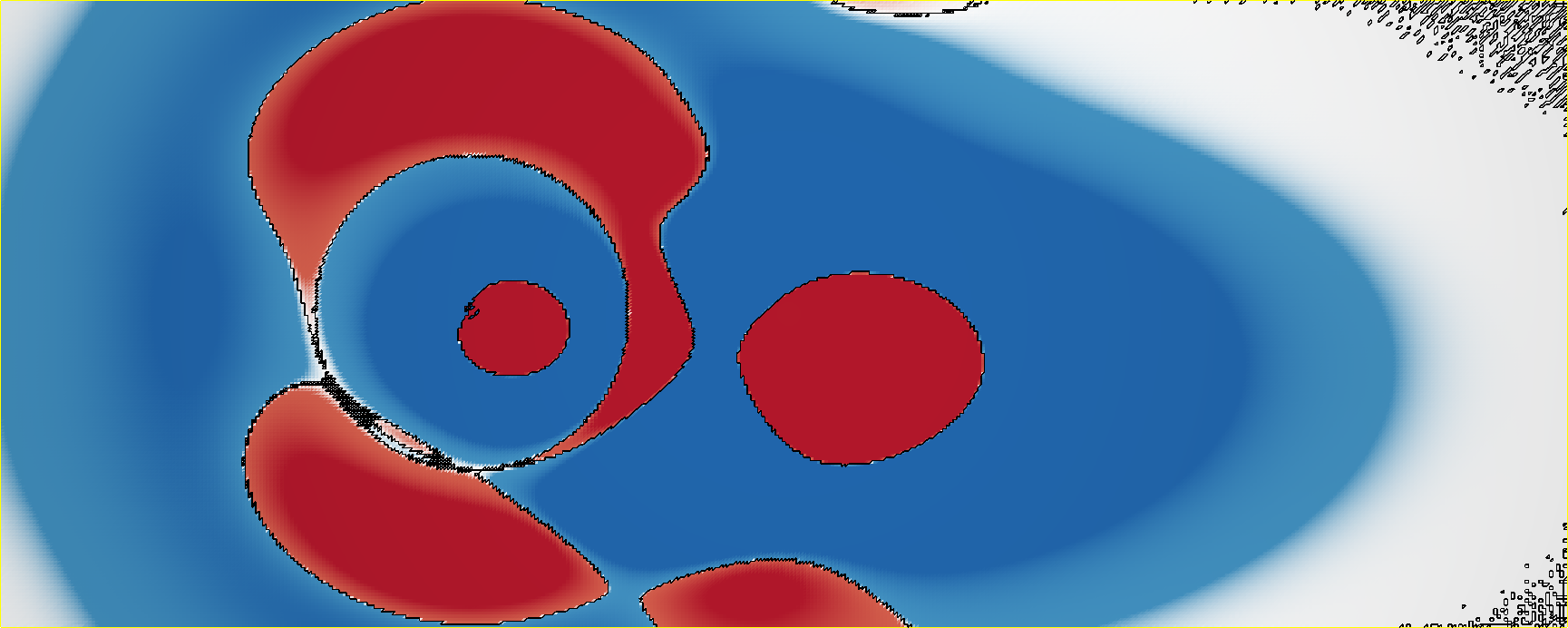}};
        \draw[Gold,ultra thick,rounded corners,rotate around={-30:(4.5,2.4)}] (3.5,2) rectangle (6,2.8);
        \draw[Brown,ultra thick,rounded corners] (9.5,7.0) rectangle (11.7,7.4);
        \draw[Green,ultra thick,rounded corners] (16.5,0.0) rectangle (18.2,1);
        \draw[Green,ultra thick,rounded corners] (15.5,6) rectangle (18.2,7.4);
    \end{tikzpicture}
    }
    \hfill
    \includegraphics[height=0.25\linewidth, alt={Nice Example to jacobi sets simplification before and after this algorithm.}]{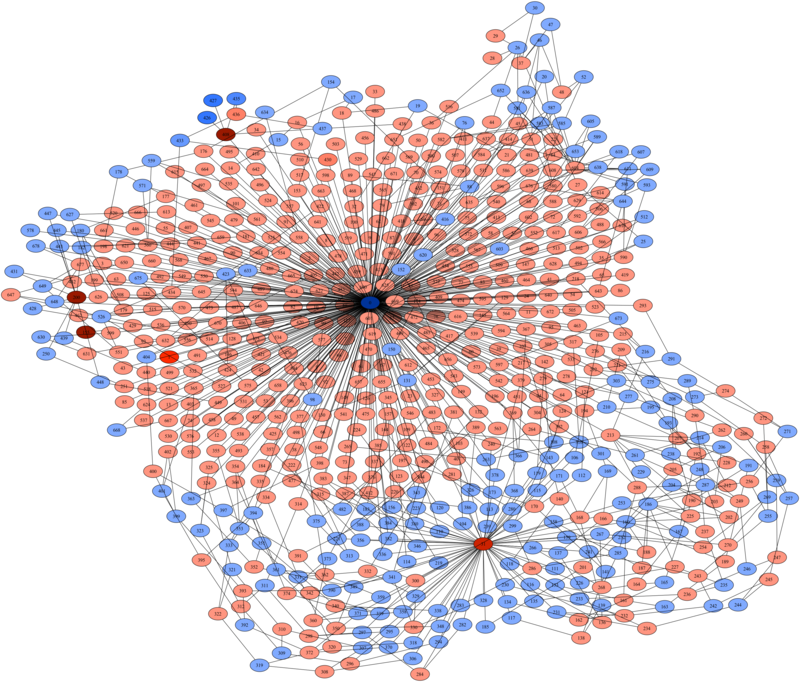}
    \resizebox{!}{0.25\linewidth}{%
    \begin{tikzpicture}
        \node[anchor=south west,inner sep=0] at (0,0)    {\includegraphics[height=0.5\linewidth, alt={Nice Example to jacobi sets simplification before and after this algorithm.}]{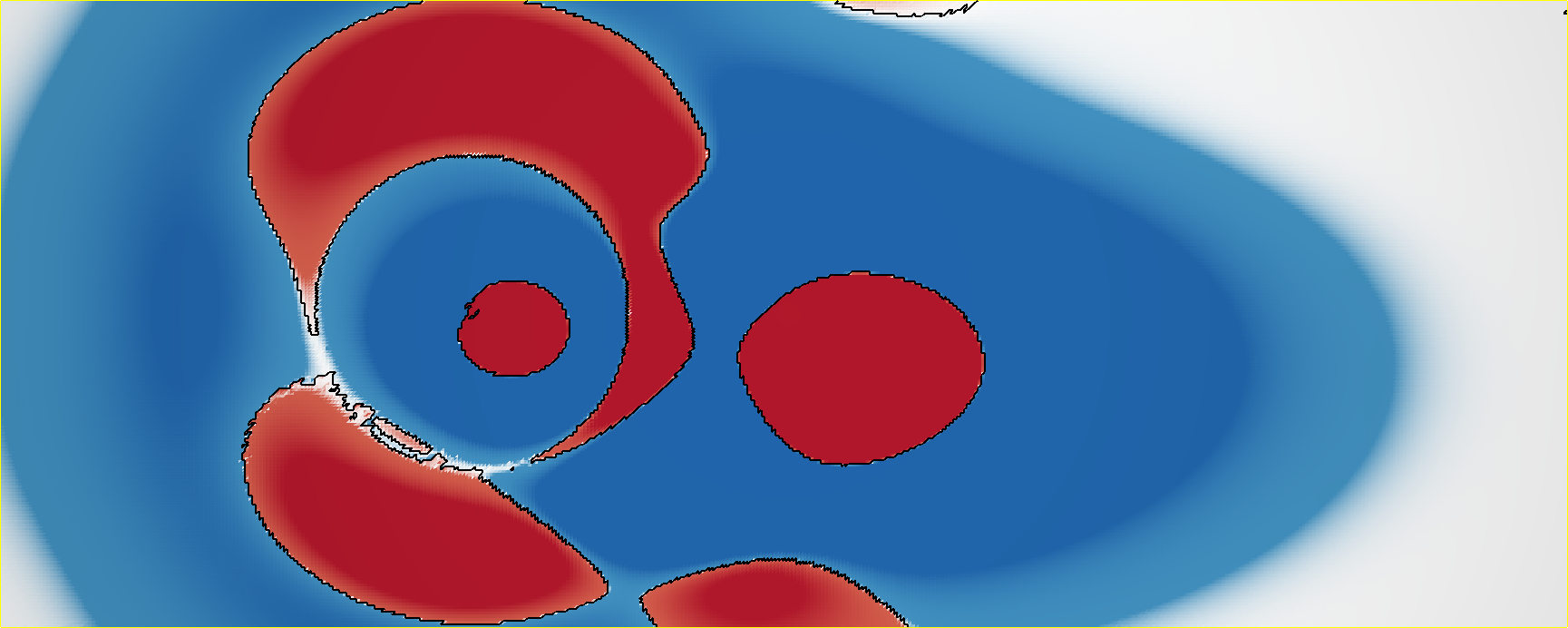}};
        \draw[Gold,ultra thick,rounded corners,rotate around={-30:(4.5,2.4)}] (3.5,2) rectangle (6,2.8);
        \draw[Brown,ultra thick,rounded corners] (9.5,7.0) rectangle (11.7,7.4);
        \draw[Green,ultra thick,rounded corners] (16.5,0.0) rectangle (18.2,1);
        \draw[Green,ultra thick,rounded corners] (15.5,6) rectangle (18.2,7.4);    \end{tikzpicture}
    }
    \hfill
    \resizebox{!}{0.25\linewidth}{%
    \includegraphics[height=0.25\linewidth, alt={Comparison of the calculated Jacobi sets in the Cylinder Flow dataset on the left side of the figure for the original dataset in the upper figure and the Collapse Algorithm Variant A in the lower figure. Furthermore, the corresponding neighborhood graphs are shown on the right side.}]{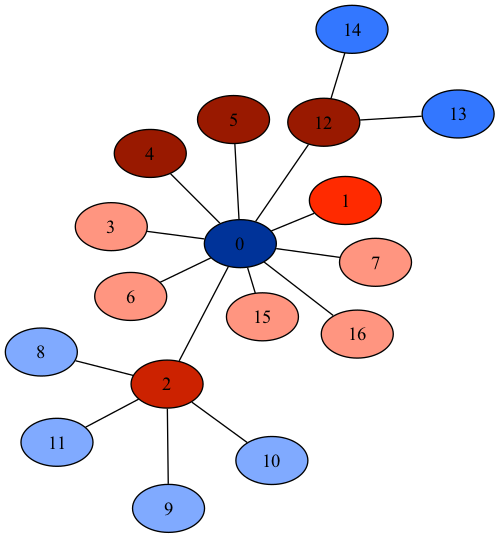}
    }
    \end{subfigure}%
    \caption{%
        Comparison of the calculated Jacobi sets in the Cylinder Flow dataset on the left side of the figure for the original dataset in the upper figure before simplification and the dataset in the lower figure after simplification with the collapse algorithm variant A with threshold value $t = 0.0001$. Furthermore, the corresponding neighborhood graphs are shown on the right side, with the color corresponding to the orientation and the range area.
    }
    \label{fig:teaser}
}

\abstract{
Jacobi sets are an important tool to study the relationship between functions.
\CH{Defined as the set of all points where the function's gradients are linearly dependent, Jacobi sets extend the notion of critical point to multifields.}
\CH{In practice, Jacobi sets for piecewise-linear approximations of smooth functions can become very complex and large due to noise and numerical errors.
Existing methods that simplify Jacobi sets exist, but either do not address how the functions' values have to change in order  to have simpler Jacobi sets or remain purely theoretical.
In this paper, we present a method that modifies 2D bivariate scalar fields such that Jacobi set components that are due to noise are removed, while preserving the essential structures of the fields.
The method uses the Jacobi set to decompose the domain, stores the and weighs the resulting regions in a neighborhood graph, which is then used to determine which regions to join by collapsing the image of the region's cells.
We investigate the influence of different tie-breaks when building the neighborhood graphs and the treatment of collapsed cells.
We apply our algorithm to a range of data sets, both analytical and real-world and compare its performance to simple data smoothing.}
} 


\newcommand{\CH}[1]{{\color{black}#1}}

\def\imageVersion{2}

\newcommand{\captionHI}{Comparison of the calculated Jacobi sets in the Hurricane Isabel dataset for the original dataset (c), the Collapse Algorithm (CA) Variant A with $t = 200$ (d), the Loop Subdivision (i), and the Gaussian filter (j). Additional comparison of the four methods in detail for cutout 1 (coordinates (772, 683) -- (1278, 1122)) (a) (e) (g) (k) and cutout 2 (coordinates (1572, 1382) -- (2077, 1822)) (b) (f) (h) (l).}

\begin{document}




\firstsection{Introduction} 

\maketitle

The analysis of multivariate data is frequently carried out in science, and the analysis of bivariate data in particular has established itself as an important field, for example in climate simulations where temperature and pressure are jointly investigated. 
The relationship between the scalar fields can be examined using Jacobi sets \cite{edelsbrunner2002jacobi} as a tool for analysis. 
This mathematical concept originates from the field of topology and represents a generalization of critical points to multifields. 
In topological data analysis, Jacobi sets are used to extract Ridge-Valley graph in computer vision and image processing \cite{norgard2013ridge}, and to track critical points in time \cite{bremer2007topological,edelsbrunner2002jacobi}. 
Besides, they are used to compare scalar fields \cite{edelsbrunner2004local} to study the Reeb space \cite{tierny2017} and to drive fiber surface extraction \cite{sharma2022jacobi}. 
They are not only used in visualization but also in other scientific disciplines \cite{artamonova2017gradient,barnett2001detection,iuricich2016discrete,makela2020automatic}. 
This multifaceted application shows the importance of Jacobi sets in scientific analysis, which is why these structures must be easy to understand.

However, noise can be a major problem that makes analysis difficult for domain experts. 
For this reason, a work on the simplification of Jacobi sets \cite{suthambhara2009simplification} has already been carried out, the Reeb graphs created and the Jacobi sets simplified with the comparison measure $\kappa$.
This involves removing loops or Jacobi set components, which occur in particular in the case of noise and numerical errors. 
However, only the representation of the Jacobi sets is simplified, the underlying data itself is not adjusted, which is unfavorable for further processing. 
In contrast, smoothing filters adjust the data globally, but the extent to which important structures are lost in the process has not been sufficiently investigated.
Important structures are regions in the dataset that lie above a threshold value of a comparison metric and are also visually separated from the surrounding area , see \cref{fig:teaser} in the upper left part in the brown region of interest.
Subdivision algorithms can also simplify Jacobi sets and, in particular, remove zig-zag, but only lead to a better representation of this \cite{klotzl2022local}.
However, the visual evaluation indicates that the number of Jacobi set components of the dataset increases.
Therefore, we will compare these approaches and our approach to determine how suitable they are for simplifying the Jacobi sets.

Motivated by the theoretical approach of Bhatia et al. \cite{bhatia2015local}, we have developed a method that aims to change the functions by \textit{collapsing} a cell to reduce the complexity of the Jacobi sets while retaining their important structures by changing the underlying data. 
The \textit{collapse} of a cell means that the area that this cell sets up in the range is reduced to the value $0$, which is also the case with degenerated cells. 
This approach simplifies the Jacobi sets and thus also reduces the Jacobi set components. 
The idea of collapsing comes from the 1D case, in which 2 critical points can be reduced by mapping the function values $f(x)$ of the two points of a cell to the same value. 
In \cref{fig:collapse1D} this is illustrated for the cell $BC$. 
Here you can see that this change affects at least one neighboring cell. 
Therefore, the choice of a suitable value is also crucial to keep the side effects as low as possible, as otherwise not all critical points can be removed. 
In the 2D case, this concept is more difficult, as the number of affected neighboring cells increases. 
Therefore, a suitable metric must be used to keep the side effects as low as possible.
To identify the components to be collapsed, we use the neighborhood graph of the Jacobi set components. 

The contributions in this paper are:
\begin{itemize}
    \item We present an algorithm for simplifying Jacobi sets based on iteratively collapsing cells while changing the underlying data and identifying this with the help of a neighborhood graph.
    \item We investigate the influence of different neighborhood graphs on this algorithm.
    \item We present an adapted Jacobi sets visualization that assigns degenerate cells according to their point neighborhood.
    \item We demonstrate the effect of smoothing and loop subdivision on the simplification of Jacobi sets using analytical and real datasets and compare them with our algorithm.
\end{itemize}


\section{Related Work} 
An overview of topology-based methods in visualization can be found in the survey by Heine et al. \cite{heine2016survey}. 
He et al. \cite{he2019multivariate} focuses more on the visualization of multivariate data.

In scalar field topology, the essential features of scalar fields can be described by contour trees \cite{carr2003computing,carr2010flexible,van1997contour}, Reeb graphs \cite{reeb1946points} or the Morse-Smale complex \cite{edelsbrunner2003hierarchical}. 
With their method, scalar fields can be topologically simplified and the critical points and their relationships can be reduced.
Carr et al. \cite{carr2004simplifying} describe a method for simplifying the contour tree by suppressing smaller topological features. 
Bremer et al. \cite{bremer2004topological} create the Morse-Smale complex of a 2D function and simplify the topology by canceling pairs of critical points. Luo et al. \cite{luo2009approximating} simplify critical points based on a point cloud. 
Edelsbrunner et al. \cite{edelsbrunner2002topological} introduce the idea of persistent homology for the topological simplification of a point cloud, which was introduced by Cohen-Steiner et al. \cite{cohen2005stability}. 
Edelsbrunner et al. \cite{edelsbrunner2002topological} introduce the idea of persistent homology for the topological simplification of a point cloud, which extended Cohen-Steiner et al. \cite{cohen2005stability} to persistence diagrams.

In order to apply the methods mentioned to multifield data, an adaptation is necessary.
Methods such as Jacobi sets and Reeb graphs cannot extended so easily.
Besides, Carlsson et al. \cite{carlsson2007theory} showed that a generalization of persistent homology is difficult. 
Singh et al. \cite{singh2007topological} use partial clustering of high-dimensional data and introduce the idea of the mapper based on this. 
The Joint Contour Net by Carr et al. \cite{carr2013joint} is partly based on this idea, but use joint contour plates and their topological connectivity. 
Reeb graphs were introduced in the works \cite{chattopadhyay2014extracting} and \cite{strodthoff2015layered} parallel for multifields. 
Chattopadhyay et al. \cite{chattopadhyay2014extracting} introduce the Jacobi structure for subdividing the Reeb space, which creates a Reeb skeleton that corresponds to the Reeb graph. 
An algorithm for calculating the Reeb space of a bivariate, piecewise-linear scalar function on a tetrahedral grid is presented by Tierny et al. \cite{tierny2016jacobi}.
Chattopadhyay et al. \cite{chattopadhyay2016multivariate} use the Joint Contour Net to further simplify multivariate data and to simplify the Reeb skeleton.

The simplification of Jacobi sets of multifield data has not seen much attention so far. 
The work by Bremer et al. \cite{bremer2007topological} describes a method that removes noise from the Jacobi sets of time-varying data. 
In the work of Hüttenberger et al. \cite{huettenberger2013towards,huettenberger2014decomposition}, a method for extending topological structures from single scalar fields to multifields using Pareto sets is proposed and the approach for simplifying Jacobi sets \cite{suthambhara2009simplification} is extended to multivariate data.
Suthambhara and Natarajan~\cite{suthambhara2009simplification} use reeb graphs to simplify the calculated Jacobi sets.
Bhatia et al. ~\cite{bhatia2015local} presented a theoretical approach to simplify Jacobi sets.
In contrast to these works, our method also aims to simplify the underlying data with a simple approach.

Another approach to simplify the data is to use non-topological methods based on smoothing.
This allows noise to be removed from the entire data, but this is heavily dependent on the filters used.
Well-known filters are the binomial and Gaussian filters. 
Tong et al. \cite{tong2003discrete} decompose the field into 3 parts, smooth them individually, and then sum them up again. 
In contrast, our method only processes the data in areas where noise occurs, which prevents other structures from being altered.

A completely different approach is presented by Klötzel et al. \cite{klotzl2022local}, they introduce a new method for calculating the Jacobi sets based on local bilinear interpolation, which implements a generalization of the definition by Edelsbrunner and Harer \cite{edelsbrunner2002jacobi}. 
This method smoothes the Jacobi sets by reducing zig-zag patterns and better-resolving structures. 
This effect shows better results, especially at high resolutions. 
However, this leads to many small Jacobi set components in very noisy areas of the dataset.


\section{Background} 
A \emph{scalar field} is a function $f: \mathbb{D} \to \mathbb{C}$, smoothly mapping from a \emph{domain} $\mathbb{D}$, which is a compact $d$-manifold, to a \emph{range} $\mathbb{C} \subset \mathbb{R}$.
In this paper, we only consider domains that are subsets of $\mathbb{R}^2$.
A \emph{critical point} $\mathbf{x}$ of $f$ is a point $\mathbf{x}\in\mathbb{D}$, where the gradient, i.e. the vector of partial derivatives, of $f$ vanishes at $x$: $\nabla f(\mathbf{x}) = \mathbf{0}$.
A bivariate scalar field can be viewed as two scalar fields $f,g$ mapping from the same domain.
Edelsbrunner and Harer~\cite{edelsbrunner2002jacobi} introduced the \emph{Jacobi set} $J(f,g)$ for a pair of
functions $f,g$ as the set of points $\mathbf{x} \in \mathbb{D}$, where the gradients $\nabla f(\mathbf{x})$ and $\nabla g(\mathbf{y})$ are linearly dependent:
\begin{equation}
\label{eq:jacobi}
\begin{split}
J(f,g) := \{ \mathbf{x} \in \mathbb{D} \mid \exists \lambda \in R: \nabla f(\mathbf{x}) + \lambda \nabla g(\mathbf{x}) & = 0 \\
\text{or } \lambda \nabla f(\mathbf{x}) + \nabla g(\mathbf{x}) & = 0 \} 
\end{split}
\end{equation}
Note that the critical points of $f$ and $g$ are trivially part of their Jacobi set.
Edelsbrunner and Harer~\cite{edelsbrunner2002jacobi} showed that, if $D$ is a 2-manifold, $J(g,f)$ is generically a collection of pairwise disjoint smooth curves, free of any self-intersections.
We will call the connected components of a Jacobi set the \emph{Jacobi set components}.
Eq.~\ref{eq:jacobi} has many equivalent formulations, for example, Edelsbrunner et al.~\cite{edelsbrunner2004local} mention that for two 2D functions, the Jacobi set is the set of all points where the rank of the Jacobian, i.e., the matrix comprised of the function's gradients, is smaller than 2.
They also present a more general criterion: the set of all points $x$ where the local $\kappa$ measure is $0$, where $\kappa(x)$ is defined as the length of the wedge product of the functions gradients.
They argue that the average of this measure over an area is related to the amount of effort needed to change the topology of the functions in that area.

A \emph{$k$-simplex} $\sigma$ is the convex hull of $k+1$ affinely independent points $P_\sigma = \{\mathbf{x}_0,\ldots,\mathbf{x}_k\}$.
2-simplices are triangles, 1-simplices are line segments and 0-simplices are points.
A simplex $\tau$ is a \emph{face} of a simplex $\sigma$ if $P_\tau \subseteq P_\sigma$.
A \emph{simplicial complex} is a set of simplices that are closed under the face relation and any two simplices' intersection is either empty or a common face.
A \emph{piecewise-linear} function assigns a function value to each 0-simplex and extends this to the other simplices using barycentric interpolation.
Edelsbrunner and Harer~\cite{edelsbrunner2002jacobi} presented an algorithm to compute the Jacobi set for piecewise-linear functions approximating smooth functions, which we will not repeat here because our algorithm works slightly simpler, due to a more restricted setting.


\section{Simplification of Bivariate 2D Scalar Fields} 
\label{sec:Approach}

Our algorithm assumes that the data are given as two piecewise linear functions $f,g$ on a compact domain $\mathbb{D} \subseteq \mathbb{R}^2$, which we will treat as a vector-valued function $\mathbf{f}: \mathbb{D} \to \mathbb{R}^2$.
It first computes a measure for each triangle and uses it to determine the Jacobi set.
It then uses the Jacobi set to decompose the domain into a set of regions separated by Jacobi set components and determines a neighborhood graph where the nodes represent the regions and are connected by an edge if separated by the same Jacobi set.
It then assigns a weight to each region by suitably aggregating the measures from the first step over all triangles of this region.
The weights are used to determine which region's functions values are to be manipulated in order to remove a Jacobi set component.

\subsection{Jacobi Set Computation}
\label{sec:Jacobi}

If $f,g$ were smooth, then due to fundamental theorems in linear algebra that relate the determinant of a matrix to the linear dependency of its rows and columns, Eq.~\ref{eq:jacobi} is equivalent to:
\begin{equation}
\label{eq:det}
J(\mathbf{f}) = \left\{ \mathbf{x} \in \mathbb{D} \mid \det \nabla \mathbf{f}(\mathbf{x}) = 0 \right\}.
\end{equation}
Note that, $\nabla \mathbf{f}(\mathbf{x})$ is called the \emph{Jacobian} of $\mathbf{f}$ in calculus, and is a matrix composed of the gradients of $f$ and $g$.
The absolute value of the Jacobian's determinant would equal the value of $\kappa$ by Edelsbrunner et al.~\cite{edelsbrunner2004local}, but by not taking the absolute value, we get an alternative method to compute the Jacobi set for piecewise-linear representations of bivariate 2D fields.
For piecewise-linear data the function restricted to the interior of each triangle $\sigma$ can be given as:
\begin{equation}
\label{eq:transform}
\mathbf{f}_{|\sigma}(\mathbf{x}) = \mathbf{A}_{\sigma}\mathbf{x}+\mathbf{b}_{\sigma}.
\end{equation}
If one uses a Taylor expansion of $\mathbf{f}_{|\sigma}$ at $\mathbf{x}_0 = 0$, one will find that $\mathbf{A}_{\sigma} = \nabla \mathbf{f}_{|\sigma}(\mathbf{x})$, and hence that the determinant of the Jacobian is constant across the triangle, but is typically discontinuous between neighboring triangles.
Nethertheless, some cells will have a positive and some a negative determinant, and we determine the Jacobi set as the list of edges which are faces to triangles with different signs of the determinant. 

Eq.~\ref{eq:transform} also gives a more geometrical intuition useful to understand the mechanics of our simplification algorithm.
Essentially, Eq.~\ref{eq:transform} can be interpreted as a linear transformation of a space, i.e., how $\mathbf{f}$ maps the triangle from the domain to the range.
Due to this transformation triangles might get translated, rotated, scaled, and mirrored.
If the determinant of $\mathbf{A}_\sigma$, i.e. the Jacobian, is negative, the triangle's image is mirrored, i.e., the order of the triangle's vertices switches between clockwise and counter-clockwise.
If the determinant is $0$ then the transformation will \emph{collapse} the triangle, i.e., its image under $\mathbf{f}$ is a line segment or a point.
We will refer to the sign of the determinant of the Jacobian as the \emph{orientation} of a triangle.
The magnitude of the determinant furthermore gives the factor by which the area of the triangle grows or shrinks, indicating how much effort it takes to change the function values at a triangle's vertices to collapse it.

\subsection{Neighborhood graph}
\label{sec:NeighborhoodGraph}
\begin{figure}[tb]
    \centering
    \begin{subfigure}[b]{\columnwidth}
        \centering
        \includegraphics[width=0.9\textwidth, alt={Example neighborhood graph a ...}]{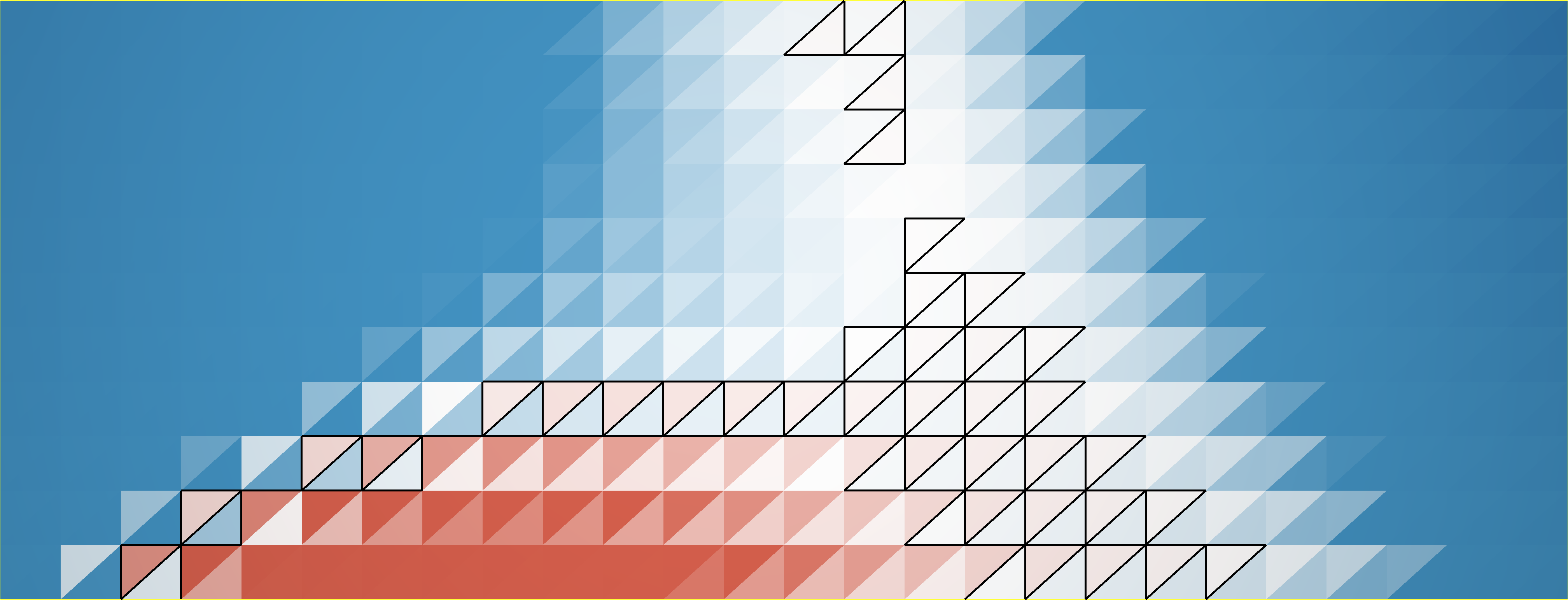}
        \caption{Cutout coordinates (-1.27, -0.52) -- (-0.67, -0.28)}
        \label{fig:ngWZoom}
    \end{subfigure}%
    \\
    \begin{subfigure}[b]{0.25\columnwidth}
        \centering
        \includegraphics[height=0.99\textwidth, alt={Example neighborhood graph a ...}]{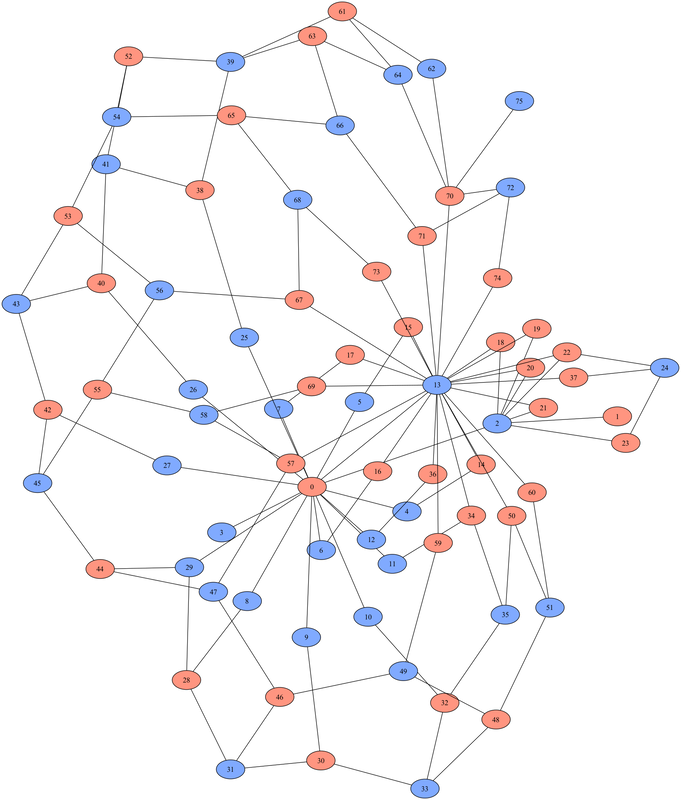}
        \caption{Neighborhood graph A}
        \label{fig:ngW}
    \end{subfigure}%
    \hfill%
    \begin{subfigure}[b]{0.25\columnwidth}
        \centering
        \includegraphics[height=0.99\textwidth, alt={Example neighborhood graph b ...}]{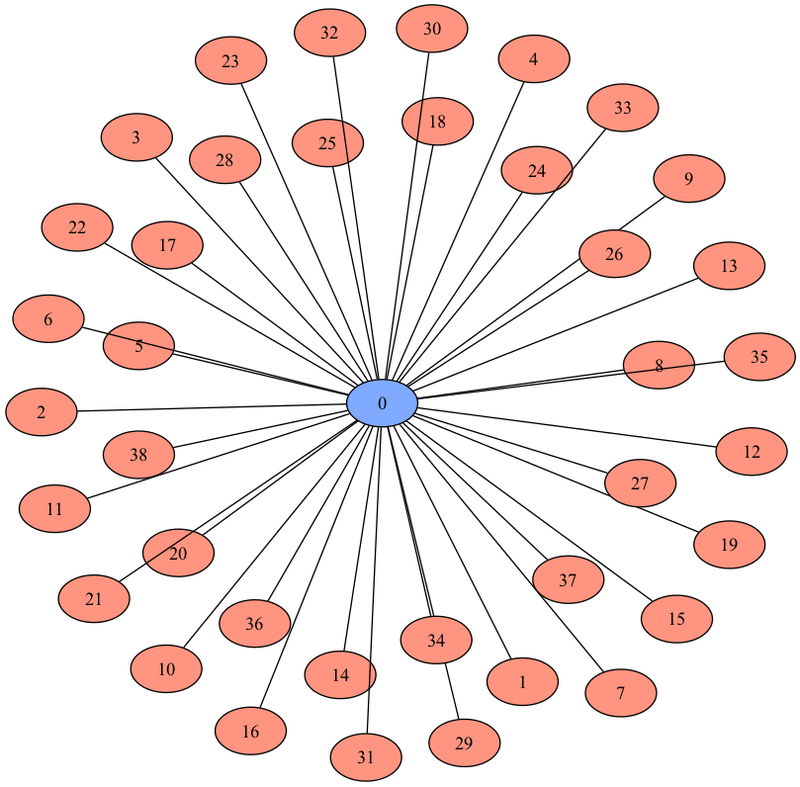}
        \caption{Neighborhood graph B}
        \label{fig:ngN}
    \end{subfigure}%
    \hfill
    \begin{subfigure}[b]{0.25\columnwidth}
        \centering
        \includegraphics[height=0.99\textwidth, alt={Example neighborhood graph c ...}]{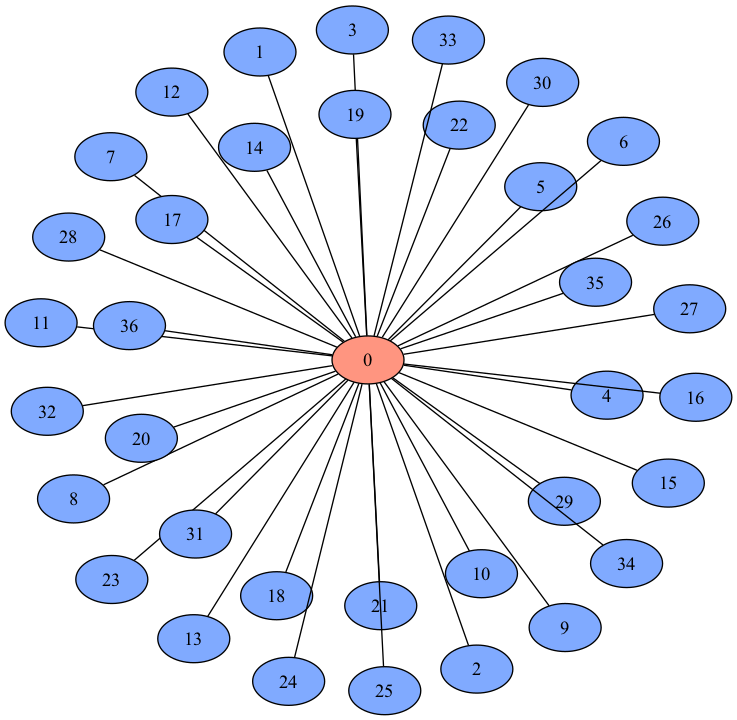}
        \caption{Neighborhood graph C}
        \label{fig:ngP}
    \end{subfigure}%
    \hfill%
    \begin{subfigure}[b]{0.25\columnwidth}
        \centering
        \includegraphics[height=0.99\textwidth, alt={Example neighborhood graph d ...}]{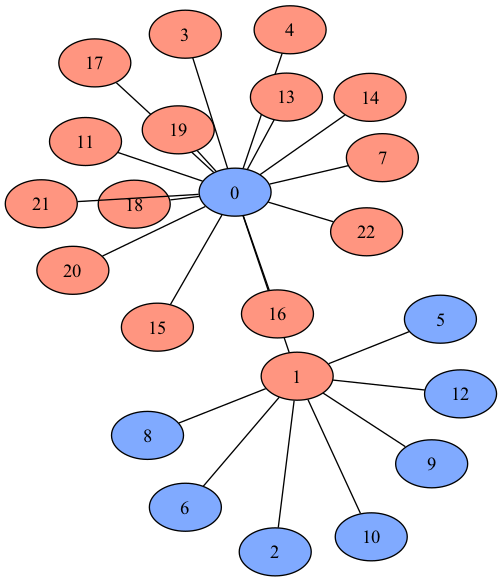}
        \caption{Neighborhood graph D}
        \label{fig:ngA}
    \end{subfigure}%
    \caption{Section of the cylinder flow dataset (a) is an example of applying the different neighborhood graphs. (b) classical neighborhood graph A, (c) neighborhood graph B with connected negative cells, (d) neighborhood graph C with connected positive cells, and (e) neighborhood graph A with connected cells corresponding to the neighborhood.}
    \label{fig:ng}
\end{figure}

A neighborhood graph represents the relationship between objects through nodes and edges.
A node represents a Jacobi set component, and an edge represents the geometric spatial proximity in the domain.
We examine four approaches for creating neighborhood graphs labeled A to D, which differ in the grouping of the cells to the components.
An example of the different of neighborhood graphs is shown in \cref{fig:ng}. 
This figure shows a section of a domain that is examined in more detail in \cref{fig:ngWZoom} and illustrates all four approaches. 
The coloring of the cells represents two properties in the range: the orientation, where positive oriented triangles are colored red and negative oriented triangles are colored blue, and the area, where high saturation means a large area.
The Jacobi sets are shown in black.

For the neighborhood graph A, all cells are combined into one node if they are not separated by a Jacobi set. 
This results in a graph with $69$ nodes and $108$ edges for the domain section under consideration.
In neighborhood graph B, as shown in \cref{fig:ngN}, cells are merged into one node if they fulfill the conditions of graph A. 
Besides, two cells are combined into one node if they are connected via a point and both have a negative orientation. 
The neighborhood graph C, shown in \cref{fig:ngP}, follows the same approach as graph B, but here two cells are combined into one node if they are connected by a point and both have a positive orientation. 
These adaptations in graphs B and C are based on the goal of converting the neighborhood graph into a neighborhood tree. 
Such a tree can be simplified by starting to combine nodes at the leaves as long as the range area of these nodes are smaller than a threshold according to a suitable metric.
A side effect of this adaptation is the reduction of nodes in the graph.
The effects of these adaptations can be seen in the resulting graphs. 
The neighborhood graph B has $39$ nodes and $38$ edges, while the graph C has $33$ nodes and $32$ edges. 
The different orientations lead to significantly different graphs.

Finally, we consider the neighborhood graph D. 
Here, all cells are combined into one node if they are not separated by a Jacobi set. 
Besides, two cells are connected if they are connected by a point, have the same orientation and the summed orientation of all point neighbors is also the same.
The goal of this approach is to minimize the imbalance that arises in the neighborhood graphs B and C and produce a more analyzable neighborhood tree. 
The resulting graph for this approach is shown in \cref{fig:ngA} and has $23$ nodes and $22$ edges. 
This graph shows a compact structure that may be well-suited for further analysis.

To investigate the effect of the different neighborhood graphs on the simplification of the Jacobi sets, all four approaches are combined with the method developed, and the results are compared visually.

\subsection{Metrics to select Jacobi set components}
\label{sec:metrics}
In this study, the goal is to collapse a selection of Jacobi set components to simplify the data. 
A suitable metric has to be chosen to select a Jacobi set component. 
There are geometric and topological metrics, and we have chosen a topological metric in this work.
Based on the neighborhood graph, we now decide which of the Jacobi set components should be collapsed. 
Different measures can be used to choose the components:

\begin{itemize}
    \item The Jacobi set component's domain area:
    \begin{equation}
    A_{\mathbb{D}} = \sum^n_{i = 0} A_i
    \end{equation}

    \item The Jacobi set components range area:
    \begin{equation}
    A'_{\mathbb{C}} = \sum^n_{i = 0} A'_i
    \end{equation}

    \item The hypervolume, which is the product of the Jacobi set components domain and range area:
    \begin{equation}
    HV = \sum^n_{i = 0} A_i \cdot A'_i
    \end{equation}
\end{itemize}

Here $i$ is a cell in a region.
We chose hypervolume, which takes into account both the area of Jacobi set components in the domain and the range.
Furthermore, the other two metrics pose problems, especially if the Jacobi set components areas are very different. 
This could lead to important structures being lost.
Another possibility would be the automatic determination of this variable depending on the neighborhood. 
However, this method was not implemented in this work.

\subsection{Collapse of Regions}
\label{sec:collapse}

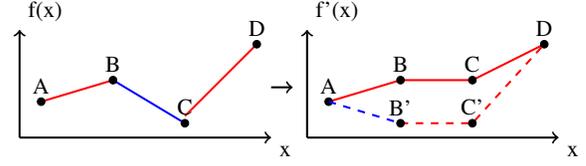
\begin{figure}[tb]
    \centering
    \resizebox{0.9\linewidth}{!}{%
    \begin{tikzpicture}
    \draw[thick,->] (0,0) -- (3.5,0) node[anchor=north west] {x};
    \draw[thick,->] (0,0) -- (0,1.5) node[anchor=south west] {f(x)};
    \draw[thick,->] (4,0) -- (7.5,0) node[anchor=north west] {x};
    \draw[thick,->] (4,0) -- (4,1.5) node[anchor=south west] {f'(x)};
    \draw[thick,->] (3.5,0.7) -- (3.8,0.7) node[anchor=north west] {};
    
    \draw [red,thick]  (0.3,0.5) -- (1.3,0.8);
    \draw [blue,thick]  (1.3,0.8) -- (2.3,0.2);
    \draw [red,thick]  (2.3,0.3) -- (3.3,1.3);
    \filldraw[black] (0.3,0.5) circle (1.5pt) node[anchor=south]{A};
    \filldraw[black] (1.3,0.8) circle (1.5pt) node[anchor=south]{B};
    \filldraw[black] (2.3,0.2) circle (1.5pt) node[anchor=south]{C};
    \filldraw[black] (3.3,1.3) circle (1.5pt) node[anchor=south]{D};
    \draw [red,thick]  (4.3,0.5) -- (5.3,0.8) -- (6.3,0.8) -- (7.3,1.3);
    \filldraw[black] (4.3,0.5) circle (1.5pt) node[anchor=south]{A};
    \filldraw[black] (5.3,0.8) circle (1.5pt) node[anchor=south]{B};
    \filldraw[black] (6.3,0.8) circle (1.5pt) node[anchor=south]{C};
    \filldraw[black] (7.3,1.3) circle (1.5pt) node[anchor=south]{D};
    \draw [blue,thick,dashed]  (4.3,0.5) -- (5.3,0.2);
    \draw [red,thick,dashed]  (5.3,0.2) -- (6.3,0.2) -- (7.3,1.3);
    \filldraw[black] (5.3,0.2) circle (1.5pt) node[anchor=south]{B'};
    \filldraw[black] (6.3,0.2) circle (1.5pt) node[anchor=south]{C'};

    \end{tikzpicture}
    }
    \caption{Example of collapsing a cell in 1D. 
Left before collapsing the cell BC, right after collapsing. 
The dashed line shows when the neighborhood is not taken into account. 
The solid line shows when it is taken into account.
The red color shows a positive and the blue a negative orientation.}
    \label{fig:collapse1D}
\end{figure}

\begin{figure}[tb]
    \centering
    \hspace*{\fill}
    \begin{subfigure}[b]{0.45\linewidth}
        \centering
        \resizebox{\textwidth}{!}{%
        \begin{tikzpicture}
        \draw[thick,->] (1.7,0.7) to [bend left] (2.5,0.8);
        \draw[white] (1.7,0.7) -- (2.5,0.8) node[black,midway, anchor=south] {V1};
        \draw[thick,->] (1.7,0.4) to [bend left] (2.5,0.3);
        \draw[white] (1.7,0.4) -- (2.5,0.3) node[black,midway, anchor=south] {V2};
        
        \draw [black,thick]  (0.3,0.2) -- (0.8,0.8) -- (1.3,0.5) -- (0.3,0.2);
        \filldraw[black] (0.3,0.2) circle (1.5pt) node[anchor=south]{A};
        \filldraw[black] (0.8,0.8) circle (1.5pt) node[anchor=south]{C};
        \filldraw[black] (1.3,0.5) circle (1.5pt) node[anchor=south]{B};
    
        \filldraw[black] (3.1,0.8) circle (1.5pt) node[anchor=west]{A'};
        \filldraw[black] (3.1,0.8) circle (1.5pt) node[anchor=south]{C'};
        \filldraw[black] (3.1,0.8) circle (1.5pt) node[anchor=east]{B'};
    
        \draw [black,thick]  (3.1,0.2) -- (4.1,0.5);
        \draw [orange,thick,dashed]  (3.1,0.2) -- (4.1,0.5);
        \filldraw[black] (3.1,0.2) circle (1.5pt) node[anchor=south]{A''};
        \filldraw[black] (4.1,0.5) circle (1.5pt) node[anchor=south]{C''};
        \filldraw[black] (4.1,0.5) circle (1.5pt) node[anchor=north]{B''};
        \end{tikzpicture}
        }
        \caption{Case 1}
        \label{fig:case1}
    \end{subfigure}%
    \hfill
    \begin{subfigure}[b]{0.45\linewidth}
        \centering
        \resizebox{\textwidth}{!}{%
        \begin{tikzpicture}
        \draw[thick,->] (1.7,0.6) to [bend left] (2.5,0.5);
        \draw[white] (1.7,0.6) -- (2.5,0.5) node[black,midway, anchor=south] {V2};
        
        \draw [black,thick]  (1.3,0.5) -- (0.3,0.2) -- (0.8,0.8);
        \draw [Green,thick,dashed]  (0.8,0.8) -- (1.3,0.5);
        \filldraw[black] (0.3,0.2) circle (1.5pt) node[anchor=south]{A};
        \filldraw[black] (0.8,0.8) circle (1.5pt) node[anchor=south]{C};
        \filldraw[black] (1.3,0.5) circle (1.5pt) node[anchor=south]{B};
    
        \draw [black,thick]  (3.1,0.2) -- (4.1,0.5);
        \draw [orange,thick,dashed]  (3.1,0.2) -- (4.1,0.5);
        \filldraw[black] (3.1,0.2) circle (1.5pt) node[anchor=south]{A''};
        \filldraw[black] (4.1,0.5) circle (1.5pt) node[anchor=south]{C''};
        \filldraw[black] (4.1,0.5) circle (1.5pt) node[anchor=north]{B''};
        \end{tikzpicture}
        }
        \caption{Case 2}
        \label{fig:case2}
    \end{subfigure}%
    \hspace*{\fill}
    \\
    \hspace*{\fill}
    \begin{subfigure}[b]{0.45\linewidth}
        \centering
        \resizebox{\textwidth}{!}{%
        \begin{tikzpicture}
        \draw[thick,->] (1.7,0.6) to [bend left] (2.5,0.5);
        \draw[white] (1.7,0.6) -- (2.5,0.5) node[black,midway, anchor=south] {V2};

        \draw [black,thick]  (1.3,0.5) -- (0.3,0.2);
        \draw [Green,thick,dashed]  (0.3,0.2) -- (0.8,0.8) -- (1.3,0.5);
        \filldraw[black] (0.3,0.2) circle (1.5pt) node[anchor=south]{A};
        \filldraw[black] (0.8,0.8) circle (1.5pt) node[anchor=south]{C};
        \filldraw[black] (1.3,0.5) circle (1.5pt) node[anchor=south]{B};
    
        \draw [black,thick]  (3.1,0.2) -- (4.1,0.5);
        \draw [orange,thick,dashed]  (3.1,0.2) -- (4.1,0.5);
        \filldraw[black] (3.1,0.2) circle (1.5pt) node[anchor=south]{A''};
        \filldraw[black] (4.1,0.5) circle (1.5pt) node[anchor=south]{C''};
        \filldraw[black] (4.1,0.5) circle (1.5pt) node[anchor=north]{B''};
        \end{tikzpicture}
        }
        \caption{Case 3}
        \label{fig:case3}
    \end{subfigure}%
    \hfill
    \begin{subfigure}[b]{0.45\linewidth}
        \centering
        \resizebox{\textwidth}{!}{%
        \begin{tikzpicture}
        \draw[thick,->] (1.7,0.6) to [bend left] (2.5,0.5);

        \draw [Green,thick,dashed]  (0.3,0.2) -- (0.8,0.8) -- (1.3,0.5) -- (0.3,0.2);
        \filldraw[black] (0.3,0.2) circle (1.5pt) node[anchor=south]{A};
        \filldraw[black] (0.8,0.8) circle (1.5pt) node[anchor=south]{C};
        \filldraw[black] (1.3,0.5) circle (1.5pt) node[anchor=south]{B};
    
        \draw [Green,thick,dashed]  (3.1,0.2) -- (3.6,0.8) -- (4.1,0.5) -- (3.1,0.2);
        \filldraw[black] (3.1,0.2) circle (1.5pt) node[anchor=south]{A};
        \filldraw[black] (3.6,0.8) circle (1.5pt) node[anchor=south]{C};
        \filldraw[black] (4.1,0.5) circle (1.5pt) node[anchor=south]{B};
        \end{tikzpicture}
        }
        \caption{Case 4}
        \label{fig:case4}
    \end{subfigure}%
    \hspace*{\fill}
    \caption{Example for the collapse of a triangle cell in the range for all 4 cases to be considered: V1 shows the collapse to a point, and V2 shows the collapse to a line. 
The green dashed edges connect the cell with a cell to be collapsed. 
The orange dashed lines are the edges to be scrambled.}
    \label{fig:collapse2DAllCases}
\end{figure}
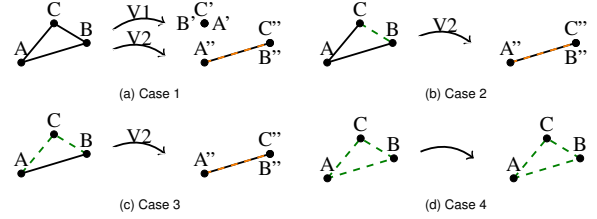

To collapse a region, we use a greedy algorithm that shrinks the regions starting from the cells at the Jacobi set and collapses them first until the entire region is collapsed.
The collapse a cell in the 2D case is a complex problem due to its influence on the neighborhood. 
In the 1D case, all values at the vertices of a cell can simply be set to the same value in the range, see \cref{fig:collapse1D}. 
This collapses the line cell to a point in the 1D case. 
In the 2D case, \cref{fig:case1} shows that $V1$ maps all values at the vertices of a triangle cell to one value and thus collapses the cell in the range.
in \cref{fig:case1} $V2$ collapsed the cell to a line in the range.
The difference between these two variants is that collapsing to a point affects more neighboring cells compared to collapsing to a line. 
Therefore, we prefer collapsing to a line in this work. 
Since only one edge of the triangle cell needs to collapse to collapse the entire cell, we decide which of the edges to select depending on the neighborhood.
In this way, side effects can be reduced, and places where several small Jacobi set components will be removed are collapsed from the outside.
Since we know at this point which cells should be collapsed, we check whether there are cells in the neighborhood that should also be collapsed, which we can then divide into four cases:
\begin{enumerate}
    \item No cell to be collapsed in the neighborhood \cref{fig:case1},
    \item One cell to be collapsed in the neighborhood \cref{fig:case2},
    \item Two cells to be collapsed in the neighborhood \cref{fig:case3},
    \item Three cells to be collapsed in the neighborhood \cref{fig:case4}.
\end{enumerate}

In the first case, all edges of the triangle can be selected for collapsing. 
In the second case, the edges that connect the cell to a cell that is not to be collapsed should be retained. 
In the third case, only the edges that connect the cell to a cell to be collapsed can be collapsed. 
We skip the fourth case, as the collapsing of the neighboring cells means that after several iterations they are also on the edge of the Jacobi set component and can then be treated according to one of the other cases.

\subsection{Method}

\begin{algorithm}
\caption{Collapse Algorithm}
\label{alg:collapse}
\DontPrintSemicolon
\SetKwInOut{Input}{Input}\SetKwInOut{Output}{Output}
\Input{2D Bivariate Scalar Field $F$,\\ Threshold $t$ }
\Output{Updated 2D Bivariate Scalar Field $F$, \\Neighborhood Graph $NG$}
\Begin{
    $NG$ := $ComputeNeighborhoodGraph(F)$\; 
    $CL$ := $FindCollapsibleCells(NG, t)$\;
    \While{$CL \neq \emptyset$}{
        \For{$c \in CL$}{
            $nc$ := $CalculateCellNeighborhood(CL, c)$\; 

            \If{$nc = 0$}{ 
                $continue$\;
            }

            $CV$ := $PossibleCollapseVariants(nc)$\;

            $bcv$ := $FindBestCollapseVariant(F, CV, c)$\;

            $ApplyCollapseVariant(F, c, bcv)$\;
            \;
            $CL$ := $CL \setminus \{c\}$\;
            $CL$ := $CL \cup FlippedCellNeighbors(NG,c)$\;
        }

        \If{$CellsOscillated(CL)$}{
            $break$\;
        }
    }
}
\end{algorithm}

The entire method is presented as pseudocode in \cref{alg:collapse}. 
Our algorithm uses a 2D bivariate scalar field defined on an unstructured grid.
For these, a neighborhood graph is created in \texttt{ComputeNeighborhoodGraph(F)}.
In this graph, the nodes represent individual Jacobi set components, while the edges represent the relationships between the different components. 
There are various ways in which the Jacobi set components can be composed in the neighborhood graph. 
These ways and their specifics are described in detail in \cref{sec:NeighborhoodGraph}.

This neighborhood graph is used in our method to select Jacobi set components. 
This is done in \texttt{FindCollapsibleCells(NG,t)}, which uses the hypervolume from \cref{sec:metrics} and an appropriate threshold to identify nodes that are considered irrelevant. 
The algorithm is cell-based and therefore returns all cells in a list (\texttt{CL}) that belong to the irrelevant nodes in a list. 
The algorithm now begins to process the list \texttt{CL}. 
For each cell, it calculates whether it lies on the border of a Jacobi set component to be collapsed.
To do this, \texttt{CalculateCellNeighborhood(CL,c)} returns for a given cell how many neighboring cells are connected to it and are not contained in \texttt{CL}.
This number is used to sort cells and skip cells that are not on the border.
For the remaining cells, the list \texttt{(CV)} is returned that contains all possibility variants of how the cell can be collapsed. 
These variant are described in more detail in \cref{sec:collapse}.
This list is now used to select the best variant for collapsing the cell. 
For this purpose, a metric is used that decides which variant has the least side effects based on the direct point neighborhood of the cell. 
This metric counts how many neighboring cells are negatively affected when the different options are applied. 
A negative influence occurs when cells change their orientation in the range, as shown by the dashed line in \cref{fig:collapse1D} for the 1D case. 
Besides, the area in the range in the neighborhood is also taken into account, and changes are preferred if the total area in the range becomes smaller.
After selecting the best variant, the bivariate scalar field in \texttt{ApplyCollapseVariant(F,c,bcv)} is adjusted.
This means that the values of the cell vertices are set to the same value according to the cases from \cref{sec:collapse}.
The cell is then removed from the list \texttt{CL}. 
Since the algorithm cannot always select a variant in which no side effects occur, \texttt{FlippedCellNeighbors(NG,c)} recognizes negatively influenced cells and then adds them to the list \texttt{CL}.

To prevent the algorithm from oscillating, the \texttt{CellsOscillated(CL)} function checks this and terminates the algorithm in this case.
The result of this method is a modified bivariate scalar field whose neighborhood graph is greatly reduced and the Jacobi sets are also simplified.

\subsection{Jacobi set Visualization}
The calculation of Jacobi sets, which separate cells according to their range, poses a challenge when a cell is degenerate or collapsed, because a degenerate cell cannot be assigned to one of the two area orientations to be separated. 
A possible solution to this problem is to assign degenerate cells based on their neighborhood.

The Jacobi sets are calculated as described in \cref{sec:Jacobi}.
Besides, degenerate cells are assigned based on their point neighborhood of the Jacobi set components that predominate in the neighborhood.
This means that the cell is assigned to the orientation that is larger when the neighboring cell orientations are summed up.
If there are only degenerated cells in the direct neighborhood, the neighbors of the degenerated cells are considered recursively until a clear assignment is possible. 

When applying this method to the four possible neighborhood graphs from \cref{sec:NeighborhoodGraph}, this leads to slightly different results when dealing with degenerated cells.
For the neighborhood graph A and the neighborhood graph D, which are mapped in \cref{fig:ngW}, \ref{fig:ngA}, collapsed cells are assigned to the Jacobi set component as described. 
In neighborhood graph B, shown in \cref{fig:ngN}, degenerated cells are assigned to a negative component as long as a negative cell exists in their neighborhood regardless of what the summed neighbor cell orientation is. 
For neighborhood graph C, which is mapped in \cref{fig:ngP}, the procedure is identical to that for the neighborhood graph B, except that it applies to positive cells. 

These adjustments ensure a clear assignment of degenerated cells, which is important for further analysis and interpretation of the Jacobi sets. 
However, it cannot be guaranteed that the length of the Jacobi set is always the shortest.


\section{Results} 
\label{sec:Results}

To evaluate how much smoothing filter and loop subdivision as well as the developed collapse algorithm (CA) in all $4$ variants simplify the Jacobi sets, we apply these algorithms to three datasets and compare them with the original data.
To do this, we compare visually and with two measures whose results for all datasets in \cref{tab:all}.
These measures are the number of Jacobi set components, and the length of all Jacobi sets in the domain.
The number of Jacobi set components is a good measure to investigate a simplification since this quantity can be derived directly from the neighborhood graph and can be considered both visually and as a quantity.
However, this quantity alone could lead to misinterpretations if many Jacobi sets are connected to a large closed edge, which reduces the components but has a negative effect on the length of the Jacobi sets. 
Therefore, we also consider the length of all Jacobi sets as a second parameter to evaluate to what extent the structure could also be simplified. 

\begin{table}[!ht]
    \caption{%
        The results for the length of the Jacobi sets, and the number of Jacobi set components can be seen for all datasets and methods tested, with the bolded values being the best. CA is short for Collapse Algorithm.%
    }
    \label{tab:all}
    \scriptsize%
    \centering%
    \begin{tabu}{*{4}{c}}
        \toprule
        Dataset / & Method & Length of & \# of Jacobi \\
        Cells & & Jacobi sets & set components \\
        \midrule
        Cylinder Flow & Original Data & 92.4614 & 679 \\
        \cref{fig:teaser}, and & Binomial filter & 64.9239 & 448 \\
        \cref{fig:jz} & Gaussian filter & \textbf{40.0536} & 101 \\
        178'702 & Loop subdivision & 82.9961 & 6'311 \\
        & CA Variant A & 44.056 & \textbf{17} \\ 
        & CA Variant B & 48.8755 & 41 \\
        & CA Variant C & 50.2173 & 42 \\
        & CA Variant D & 49.8907 & 23 \\
        \midrule
        Tensile Bar A & Original Data & 1382.95 & 817 \\
        \cref{fig:tb0W} / & Loop subdivision & 1845.95 & 17'858 \\
        30'960 & CA Variant A & \textbf{615.601} & \textbf{19} \\ 
        \midrule
        Tensile Bar B & Original Data & 1476.09 & 800 \\
        \cref{fig:tb1W} / & Loop subdivision & 2200.9 & 23'625 \\
        22'360 & CA Variant A & \textbf{767.146} & \textbf{50} \\ 
        \midrule
        Tensile Bar C & Original Data & 1503.17 & 883 \\
        \cref{fig:tb2W} / & Loop subdivision & 2068.83 & 21'529 \\
        29'560 & CA Variant A & \textbf{675.679} & \textbf{18} \\ 
        \midrule
        Tensile Bar D & Original Data & 963.471 & 519 \\
        \cref{fig:tb3W} / & Loop subdivision & 1010.44 & 7'164 \\
        26'850 & CA Variant A & \textbf{465.583} & \textbf{40} \\ 
        \midrule
        Tensile Bar E & Original Data & 1004.65 & 490 \\
        \cref{fig:tb4W} / & Loop subdivision & 1238.53 & 9'442 \\
        36'336 & CA Variant A & \textbf{535.383} & \textbf{38} \\ 
        \midrule
        Tensile Bar F & Original Data & 791.891 & 329 \\
        \cref{fig:tb5W} / & Loop subdivision & 1004.82 & 7'259 \\
        27'210 & CA Variant A & \textbf{478.767} & \textbf{24} \\ 
        \midrule
        Tensile Bar G & Original Data & 815.444 & 326 \\
        \cref{fig:tb6W} / & Loop subdivision & 1094.49 & 6'827 \\
        34'012 & CA Variant A & \textbf{514.535} & \textbf{36} \\ 
        \midrule
        Tensile Bar H & Original Data & 782.771 & 322 \\
        \cref{fig:tb7W} / & Loop subdivision & 846.473 & 4'484 \\
        28'226 & CA Variant A & \textbf{447.887} & \textbf{28} \\ 
        \midrule
        Hurricane & Original Data & 915'064 & 43'838 \\
        Isabel & Binomial filter & 662'059 & 27'344 \\
        \cref{fig:hi} / & Gaussian filter & \textbf{122'871} & 4'832 \\
        498'002 & Loop subdivision & 801'508 & 336'823 \\
        & CA Variant A & 393'343 & \textbf{2'657} \\ 
        \bottomrule
    \end{tabu}%
\end{table}

In all figures, the color in the domain shows the orientation of the cell in the range, while at the same time, the range area is visualized via saturation.
The demonstrations were run on a MacBook Pro (14-inch, 2021) with a Apple M1 Max, and 64 GB Ram. 
Implemented is the algorithm inside our framework with C++.

\subsection{Simple Smoothing Approaches}

For our evaluations, we use known smoothing methods on the one hand and loop subdivision on the other.
These are classic smoothing filters to reduce noise and remove irregularities.
More precisely, we use the Gaussian filter and the binomial filter.
Loop subdivision is a method from computer graphics for smooth subdivision of triangular grids. 
This method, developed by Loop \cite{loop1987smooth}, has become a fundamental tool in computer graphics.
For all comparisons we use $\sigma = 1000$ for the Gaussian filter, $r = 1$ for the binomial filter, and $4$ subdivision steps in the loop subdivision.
In the appendix, there are further comparisons with other $\sigma$ settings for the Gaussian filter.

\subsection{Cylinder Flow (Synthetic)}
First, we consider the analytical \textit{Cylinder Flow} \cite{Jung93} dataset, which is freely available on the ETH Zürich website \cite{ethZuerich}.
The dataset is a regular 2D grid with a resolution of $450 \times 200$ and $500$ time steps, on which a synthetic vector field is defined.
This vector field was used as a co-gradient to a stream function of Jung et al. \cite{Jung93} and represents a simplified model of a K\'{a}rm\'{a}n vortex street.
We use time step $1$ for the comparisons and triangulate the grid.

\begin{figure*}[tb]
    \centering 
    \begin{subfigure}[b]{0.33\textwidth}
        \centering
        \resizebox{\textwidth}{!}{%
        \begin{tikzpicture}
            \node[anchor=south west,inner sep=0] at (0,0)    {\includegraphics[height=1.235\textwidth, alt={Nice Example to Jacobi sets simplification before and after this algorithm.}]{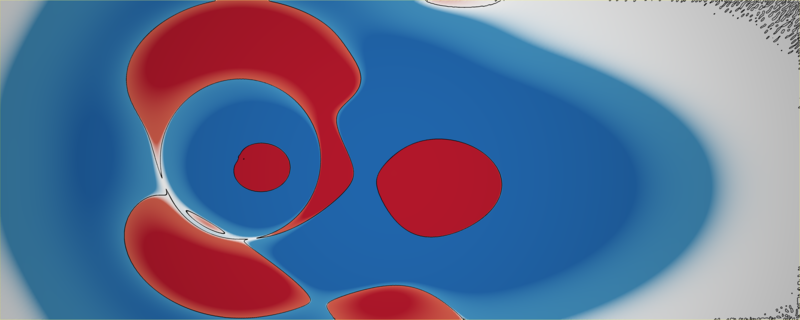}};
            \draw[Gold,ultra thick,rounded corners,rotate around={-30:(4.5,2.4)}] (3.5,2) rectangle (6,2.8);
            \draw[Brown,ultra thick,rounded corners] (9.5,7.0) rectangle (11.7,7.4);
            \draw[Green,ultra thick,rounded corners] (16.5,0.0) rectangle (18.2,1);
            \draw[Green,ultra thick,rounded corners] (15.5,6) rectangle (18.2,7.4);
        \end{tikzpicture}
        }
        \caption{Loop Subdivision}
        \label{fig:jzL4}
    \end{subfigure}%
    \hfill%
    \begin{subfigure}[b]{0.33\textwidth}
        \centering
        \resizebox{\textwidth}{!}{%
        \begin{tikzpicture}
            \node[anchor=south west,inner sep=0] at (0,0)    {\includegraphics[height=1.235\textwidth, alt={Nice Example to Jacobi sets simplification before and after this algorithm.}]{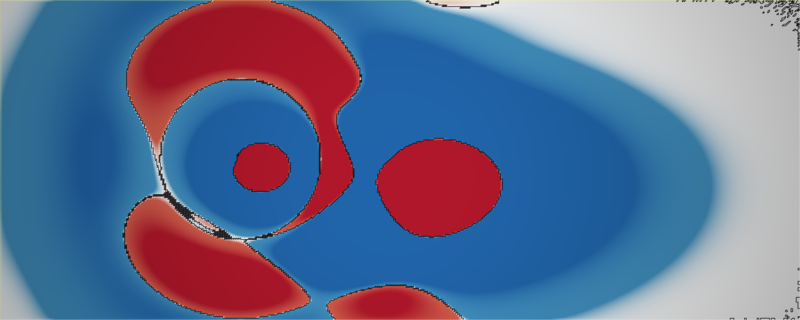}};
            \draw[Gold,ultra thick,rounded corners,rotate around={-30:(4.5,2.4)}] (3.5,2) rectangle (6,2.8);
            \draw[Brown,ultra thick,rounded corners] (9.5,7.0) rectangle (11.7,7.4);
            \draw[Green,ultra thick,rounded corners] (16.5,0.0) rectangle (18.2,1);
            \draw[Green,ultra thick,rounded corners] (15.5,6) rectangle (18.2,7.4);    
        \end{tikzpicture}
        }
        \caption{Binomial filter}
        \label{fig:jzB}
    \end{subfigure}%
    \hfill%
    \begin{subfigure}[b]{0.33\textwidth}
        \centering
        \resizebox{\textwidth}{!}{%
        \begin{tikzpicture}
            \node[anchor=south west,inner sep=0] at (0,0)    {\includegraphics[height=1.235\textwidth, alt={Nice Example to Jacobi sets simplification before and after this algorithm.}]{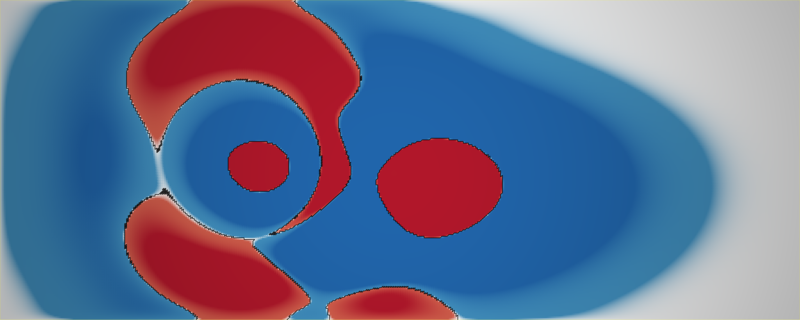}};
            \draw[Gold,ultra thick,rounded corners,rotate around={-30:(4.5,2.4)}] (3.5,2) rectangle (6,2.8);
            \draw[Brown,ultra thick,rounded corners] (9.5,7.0) rectangle (11.7,7.4);
            \draw[Green,ultra thick,rounded corners] (16.5,0.0) rectangle (18.2,1);
            \draw[Green,ultra thick,rounded corners] (15.5,6) rectangle (18.2,7.4);    
        \end{tikzpicture}
        }
        \caption{Gaussian filter}
        \label{fig:jzG}
    \end{subfigure}%
    \\
    \begin{subfigure}[b]{0.33\textwidth}
        \centering
        \resizebox{\textwidth}{!}{%
        \begin{tikzpicture}
            \node[anchor=south west,inner sep=0] at (0,0)    {\includegraphics[height=1.235\textwidth, alt={Nice Example to Jacobi sets simplification before and after this algorithm.}]{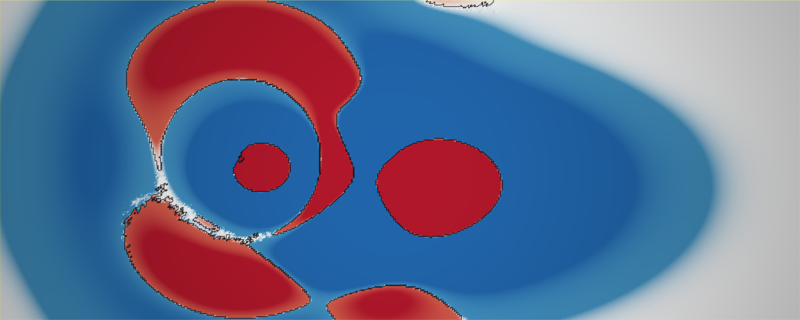}};
            \draw[Gold,ultra thick,rounded corners,rotate around={-30:(4.5,2.4)}] (3.5,2) rectangle (6,2.8);
            \draw[Brown,ultra thick,rounded corners] (9.5,7.0) rectangle (11.7,7.4);
            \draw[Green,ultra thick,rounded corners] (16.5,0.0) rectangle (18.2,1);
            \draw[Green,ultra thick,rounded corners] (15.5,6) rectangle (18.2,7.4);
        \end{tikzpicture}
        }
        \caption{Collapse Algorithm Variant B}
        \label{fig:jzN}
    \end{subfigure}%
    \hfill%
    \begin{subfigure}[b]{0.33\textwidth}
        \centering
        \resizebox{\textwidth}{!}{%
        \begin{tikzpicture}
            \node[anchor=south west,inner sep=0] at (0,0)    {\includegraphics[height=1.235\textwidth, alt={Nice Example to Jacobi sets simplification before and after this algorithm.}]{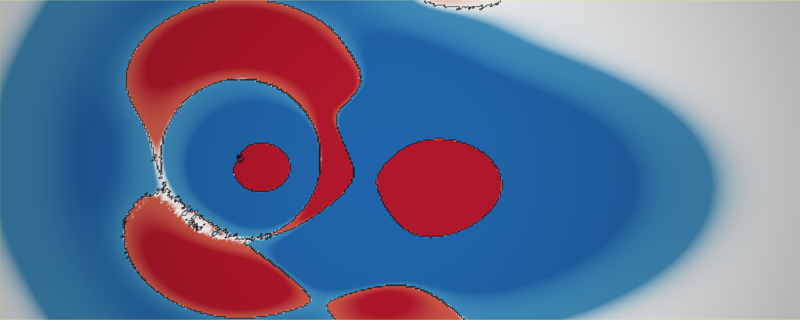}};
            \draw[Gold,ultra thick,rounded corners,rotate around={-30:(4.5,2.4)}] (3.5,2) rectangle (6,2.8);
            \draw[Brown,ultra thick,rounded corners] (9.5,7.0) rectangle (11.7,7.4);
            \draw[Green,ultra thick,rounded corners] (16.5,0.0) rectangle (18.2,1);
            \draw[Green,ultra thick,rounded corners] (15.5,6) rectangle (18.2,7.4);    
        \end{tikzpicture}
        }
        \caption{Collapse Algorithm Variant C}
        \label{fig:jzP}
    \end{subfigure}%
    \hfill%
    \begin{subfigure}[b]{0.33\textwidth}
        \centering
        \resizebox{\textwidth}{!}{%
        \begin{tikzpicture}
            \node[anchor=south west,inner sep=0] at (0,0)    {\includegraphics[height=1.235\textwidth, alt={Nice Example to Jacobi sets simplification before and after this algorithm.}]{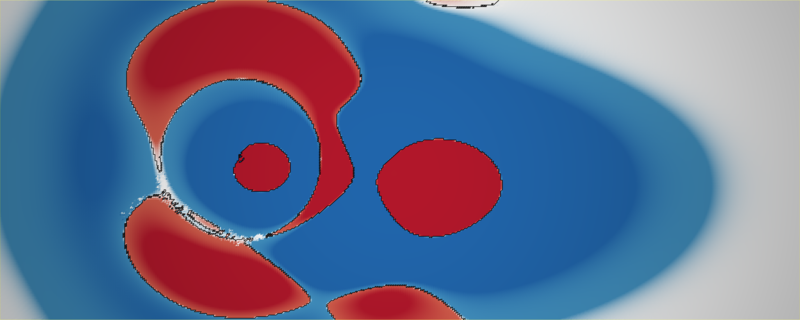}};
            \draw[Gold,ultra thick,rounded corners,rotate around={-30:(4.5,2.4)}] (3.5,2) rectangle (6,2.8);
            \draw[Brown,ultra thick,rounded corners] (9.5,7.0) rectangle (11.7,7.4);
            \draw[Green,ultra thick,rounded corners] (16.5,0.0) rectangle (18.2,1);
            \draw[Green,ultra thick,rounded corners] (15.5,6) rectangle (18.2,7.4);    
        \end{tikzpicture}
        }
        \caption{Collapse Algorithm Variant D}
        \label{fig:jzA}
    \end{subfigure}%
    \caption{Comparison of the calculated Jacobi sets in the Cylinder Flow dataset for the Loop Subdivision (a), the Binomial filter (b), the Gaussian filter (c), the Collapse Algorithm variant B with $t = 0.0001$ (d), Collapse Algorithm variant C with $t = 0.0001$ (e), and Collapse Algorithm variant D with $t = 0.0001$ (f). Three regions of interest (ROI) are highlighted in color for visual comparison.}%
    \label{fig:jz}
\end{figure*}

The aim of the analysis of this dataset is to show the effect of the CA in all variants, the smoothing filters and the loop subdivision on the simplification of the Jacobi sets and to find out which CA variant has the greatest effect and to use this for the other datasets.
The original data serves as a reference.
The resulting datasets and the corresponding Jacobi sets are shown in \cref{fig:teaser} and \cref{fig:jz}.

When observing the datasets, three regions of interest (ROI) can be identified that are relevant for the comparison because they are noisier or the structures are complex.
Nevertheless, the rest of the dataset is not uninteresting as important structures should not or only minimally be changed.
The first ROI is marked green in the datasets and contains the two right corners where many small Jacobi set components can be recognized in \cref{fig:teaser} of the original data.
These are due to noise, as the range area is very small, which can also be seen from the saturation of the colors.
In the loop subdivision, the Jacobi set components can still be recognized very well; a simplification of the Jacobi sets is therefore not possible.
After smoothing with the binomial filter, a reduction of the Jacobi set components can already be recognized. 
The smoothing is further intensified with the Gaussian filter, which even leads to the removal of all Jacobi set components in this ROI.
A similar result is obtained after applying the CA variants.
Here, all Jacobi set components can also be removed, which can be seen in \cref{fig:jzN}, e, f.
Only in variant A do individual Jacobi set components remain.
The second ROI is marked in brown in the top center of the original data.
A larger reddish Jacobi set component can be seen here, with several smaller components adjacent to it , which should be removed.
After applying the loop subdivision, the large Jacobi set component is visible and the small ones are no longer recognizable.
When smoothing with the binomial filter, there is hardly any effect in \cref{fig:jzB} and the large and small Jacobi set components are still present.
In contrast, the effect of smoothing by the Gaussian filter is too large and all components are removed, which is visible in \cref{fig:jzG}.
When using the CA variant D, the result is similar to the binomial filter, and all components are still present.
CA variants B and C do better here, where the large Jacobi set component remains and almost all small components disappear.
In CA variant A, only the large component remains, as can be seen in \cref{fig:teaser} at the bottom left.
The third ROI is marked in yellow in the center of the original data. 
In this ROI, several small Jacobi set components can be recognized in the original data.
In the loop subdivision, several large Jacobi set components are separated from each other as can be seen in \cref{fig:jzL4}.
This indicates that the small components are caused by noise or numerical errors and should disappear.
With the smoothing filters, the result is similar to before.
The binomial filter smoothes too little and the Gaussian filter too much, whereby in this ROI the Gaussian filter does not remove all the small Jacobi set components, but even creates many new ones.
Using the CA variants produces very different results here.
In variant C, for example, the Jacobi set components are not separated but connected.
In variants B and D, the larger components are separated from each other, but many small components remain.
In variant A, the large components of the Jacobi set are completely separated and almost all small components disappear so that the results of the loop subdivision are very similar.

\begin{figure}[tb]
    \centering 
    \hspace*{\fill}
    \begin{subfigure}[b]{0.45\columnwidth}
        \centering
        \includegraphics[width=\textwidth, alt={Example neighborhood graph c ...}]{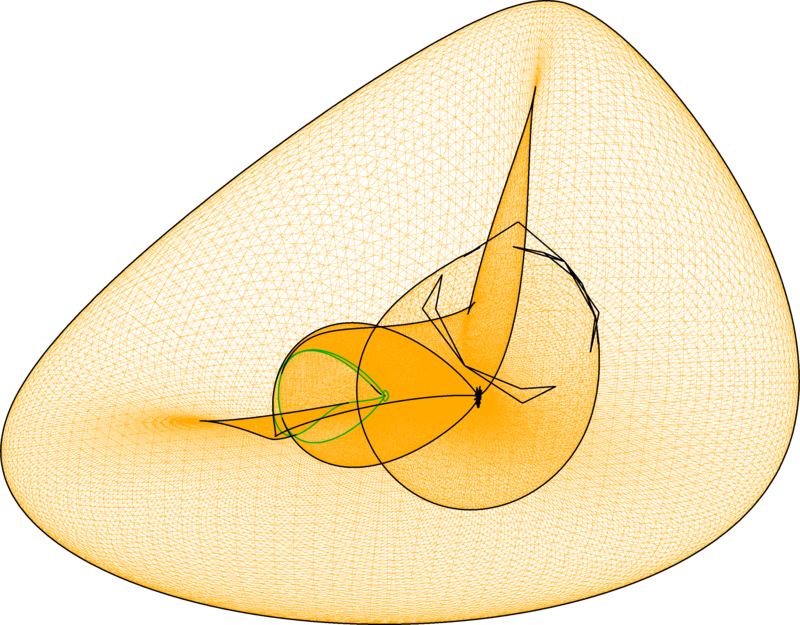}
        \caption{}
        \label{fig:jzWCodo}
    \end{subfigure}%
    \hfill%
    \begin{subfigure}[b]{0.45\columnwidth}
        \centering
        \includegraphics[width=\textwidth, alt={Example neighborhood graph d ...}]{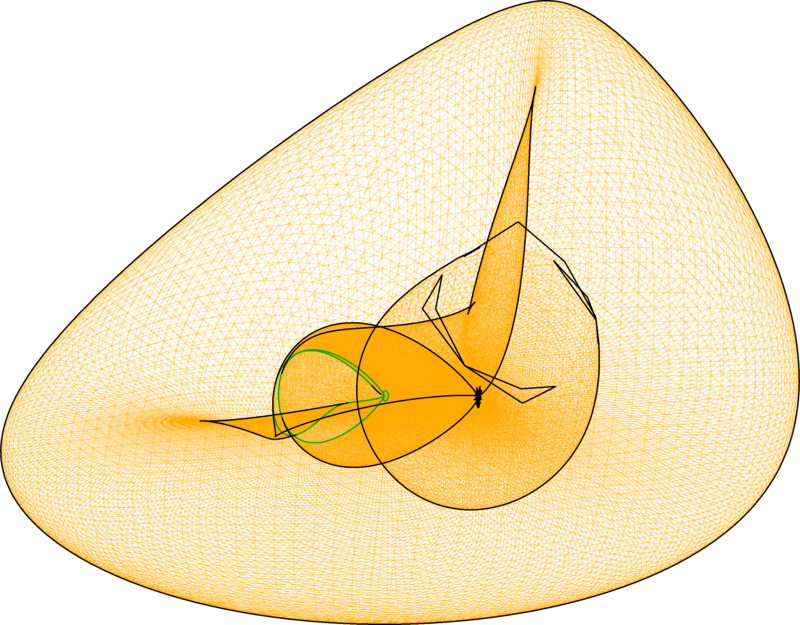}
        \caption{}
        \label{fig:jzCCodo}
    \end{subfigure}%
    \hspace*{\fill}
    \caption{The datasets from \cref{fig:teaser} are mapped into the range.
The border of the domain is shown in green, the Jacobi sets in black, and the grid in orange for the original data (a) and the dataset after applying CA variant A (b) in orange.}
    \label{fig:jzCodo}
\end{figure}

Overall, the visual comparison of the cylinder flow shows that the CA variant A is the best variant.
This is also shown in the \cref{tab:all}.
To confirm this, the \cref{fig:teaser} on the right shows the neighborhood graphs of the Jacobi set components for the original data and for the dataset after applying CA variant A. 
Here we can see that the algorithm greatly simplifies the Jacobi set components.
Besides, the range of the original data is compared with the dataset after applying the CA variant A in \cref{fig:jzCodo} to examine the effects of modifying the data.
Despite the differences in the domain, there is hardly any visual difference between the two datasets. 
The simplification from the Jacobi set components from $679$ to $17$ using the CA variant A takes $0.2$\,s.

\subsection{Tensile Bar}
\begin{figure}[tb]
    \centering
    \includegraphics[width=0.9\columnwidth, alt={The geometry of the 3D tensile bar F can be seen as an example.}]{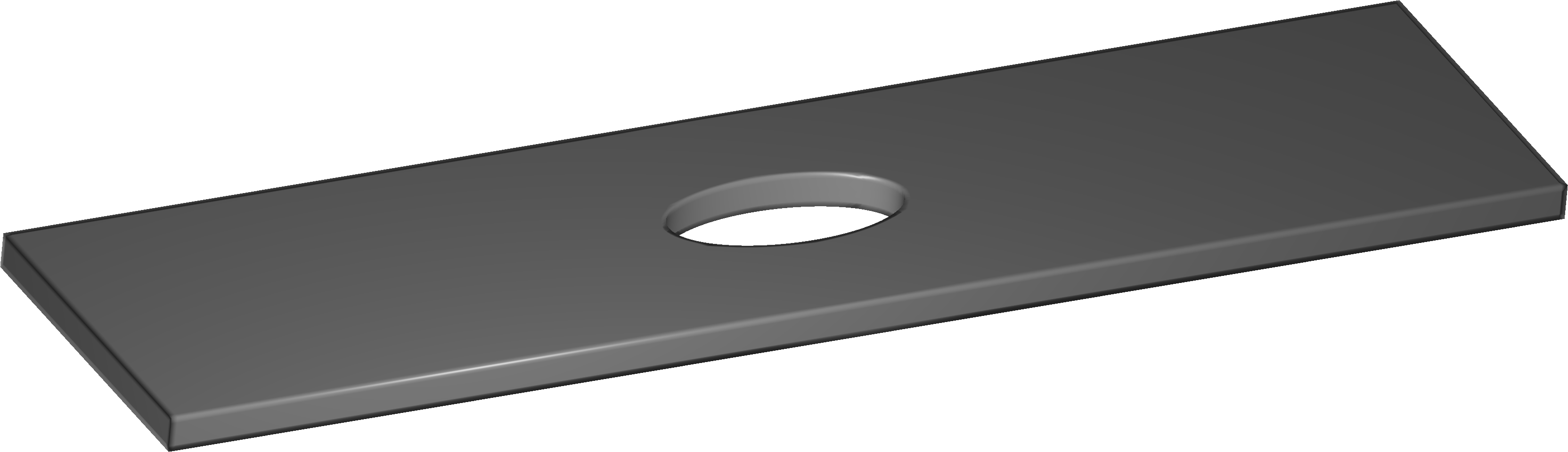}
    \caption{%
        Example of the 3D tensile bar F geometry.%
    }
    \label{fig:tb3D}
\end{figure}

Next, we look at a collection of real-world datasets, the \textit{Tensile bars} \cite{zobel2018extremal,zobel2017visualizing,raith2023fast}.
Various notches or holes are made in these specimens, as shown in \cref{fig:tb3D}, to create different loading conditions and test the material properties.
The tensile bars were simulated using the commercial software package Abaqus/Standard CAE \cite{abaqus}.
The datasets are unstructured 3D datasets on which 3D symmetric tensor fields of second order are defined.
This datasets have a dimension of the volume of the dataset of $120 \times 30 \times 3$.
Since the tensile forces are mainly in one plane, we were able to use a single layer from the datasets and triangulate the grid.
Here, the 3D tensor is mapped to a 2D tensor from which we derive an invariant set for the comparison.
For this comparison, we use the principal invariants $I$ which are the coefficients of the characteristic polynomial.
They are defined for 2D tensor fields as $I_1=trT=\lambda_1+\lambda_2$, and $I_2 =detT =\lambda_1\cdot\lambda_2$, where $trT$ denotes the trace and $detT$ the determinant of $T$ for the eigenvalues $\lambda$.
A detailed description of these and other stress tensors can be found in Holzapfel's textbook \cite{holzapfel2000nonlinear}.

\begin{figure*}[tb]
    \centering 
    \hspace*{\fill}
    \begin{subfigure}[b]{0.10\textwidth}
        \centering
        \includegraphics[width=\textwidth, alt={Jacobi sets in the original data of Tensile Bar A.}]{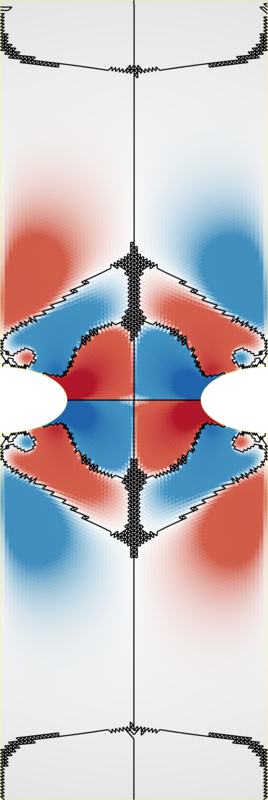}
        \caption{Tensile Bar A}
        \label{fig:tb0W}
    \end{subfigure}%
    \hfill%
    \begin{subfigure}[b]{0.075\textwidth}
        \centering
        \includegraphics[width=\textwidth, alt={Jacobi sets in the original data of tensile bar B.}]{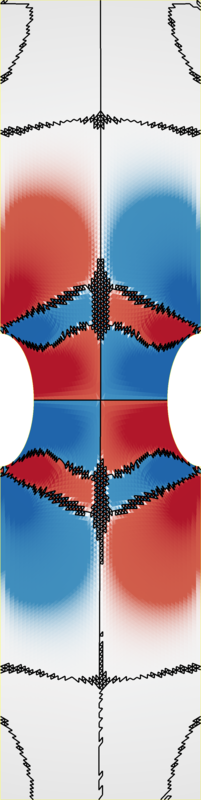}
        \caption{Tensile Bar B}
        \label{fig:tb1W}
    \end{subfigure}%
    \hfill%
    \begin{subfigure}[b]{0.10\textwidth}
        \centering
        \includegraphics[width=\textwidth, alt={Jacobi sets in the original data of tensile bar C.}]{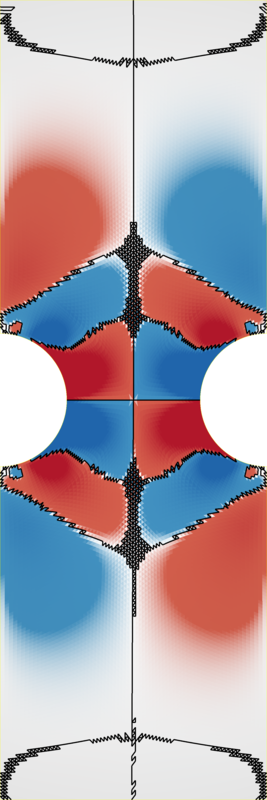}
        \caption{Tensile Bar C}
        \label{fig:tb2W}
    \end{subfigure}%
    \hfill%
    \begin{subfigure}[b]{0.075\textwidth}
        \centering
        \includegraphics[width=\textwidth, alt={Jacobi sets in the original data of tensile bar D.}]{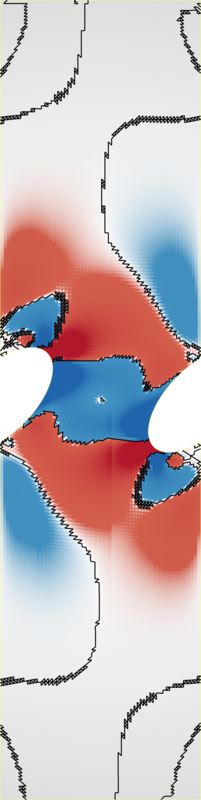}
        \caption{Tensile Bar D}
        \label{fig:tb3W}
    \end{subfigure}%
    \hfill%
    \begin{subfigure}[b]{0.10\textwidth}
        \centering
        \includegraphics[width=\textwidth, alt={Jacobi sets in the original data of tensile bar E.}]{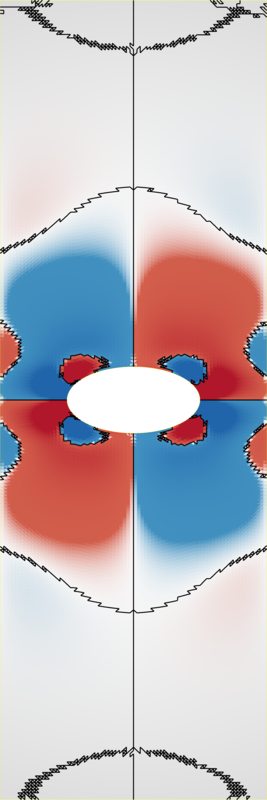}
        \caption{Tensile Bar E}
        \label{fig:tb4W}
    \end{subfigure}%
    \hfill%
    \begin{subfigure}[b]{0.075\textwidth}
        \centering
        \includegraphics[width=\textwidth, alt={Jacobi sets in the original data of tensile bar F.}]{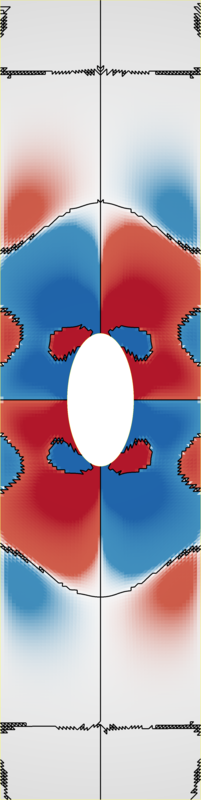}
        \caption{Tensile Bar F}
        \label{fig:tb5W}
    \end{subfigure}%
    \hfill%
    \begin{subfigure}[b]{0.10\textwidth}
        \centering
        \includegraphics[width=\textwidth, alt={Jacobi sets in the original data of tensile bar G.}]{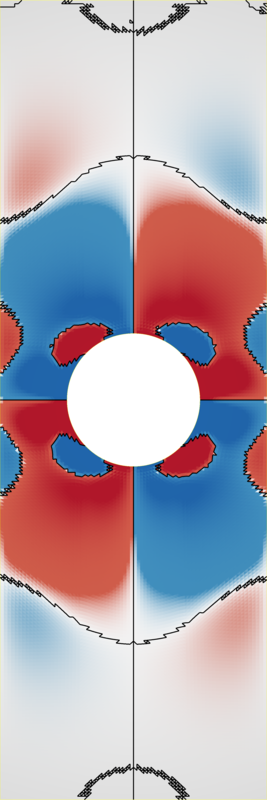}
        \caption{Tensile Bar G}
        \label{fig:tb6W}
    \end{subfigure}%
    \hfill%
    \begin{subfigure}[b]{0.075\textwidth}
        \centering
        \includegraphics[width=\textwidth, alt={Jacobi sets in the original data of tensile bar H.}]{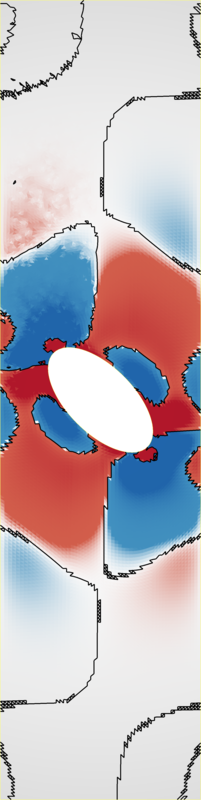}
        \caption{Tensile Bar H}
        \label{fig:tb7W}
    \end{subfigure}%
    \hspace*{\fill}
    \\
    \hspace*{\fill}
    \begin{subfigure}[b]{0.10\textwidth}
        \centering
        \includegraphics[width=\textwidth, alt={Calculated Jacobi sets with Collapse Algorithm Variant A in the tensile bar A.}]{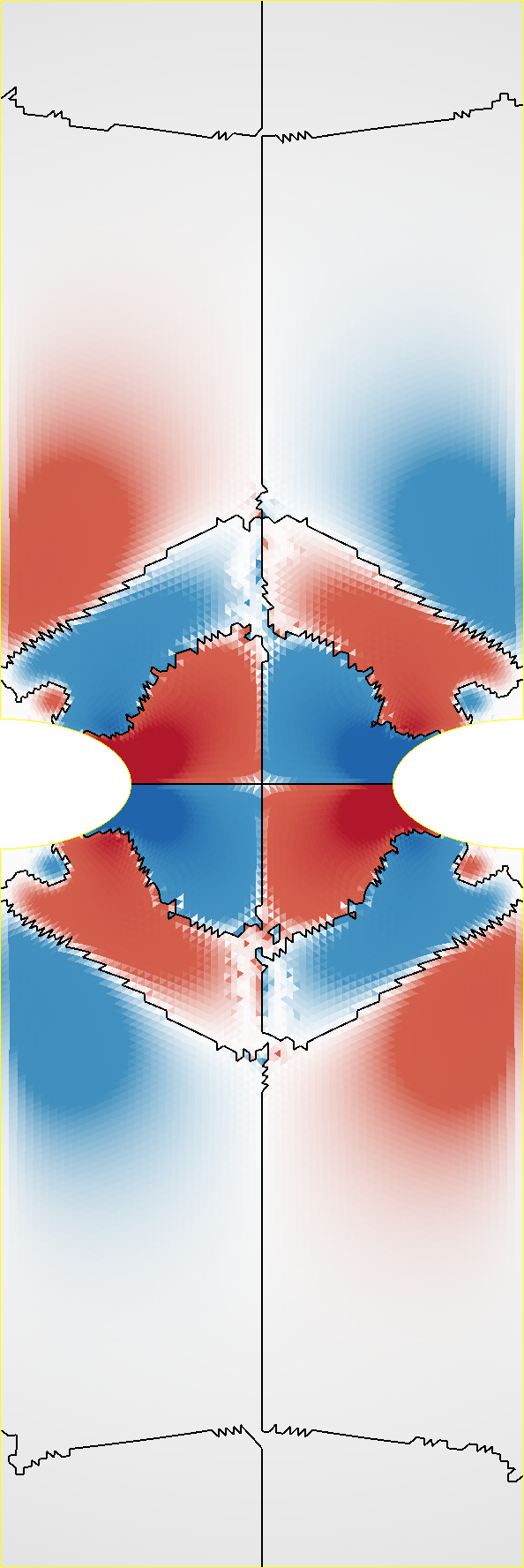}
        \caption{$t = 0.001$}
        \label{fig:tb0C}
    \end{subfigure}%
    \hfill%
    \begin{subfigure}[b]{0.075\textwidth}
        \centering
        \includegraphics[width=\textwidth, alt={Calculated Jacobi sets with Collapse Algorithm Variant A in the tensile bar B.}]{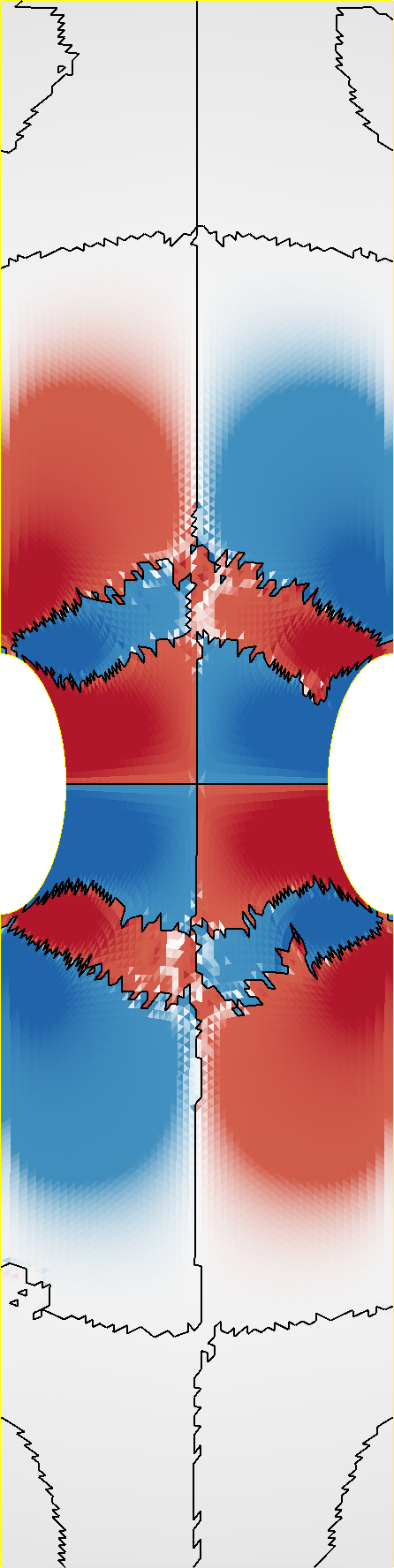}
        \caption{$t = 0.001$}
        \label{fig:tb1C}
    \end{subfigure}%
    \hfill%
    \begin{subfigure}[b]{0.10\textwidth}
        \centering
        \includegraphics[width=\textwidth, alt={Calculated Jacobi sets with Collapse Algorithm Variant A in the tensile bar C.}]{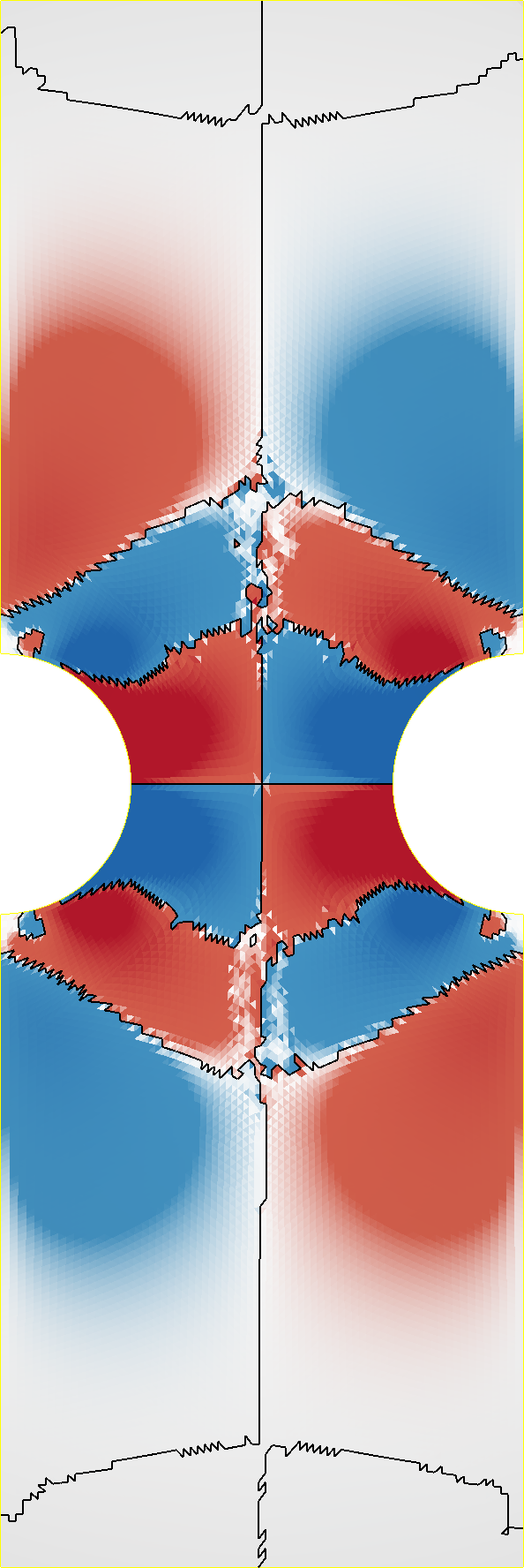}
        \caption{$t = 0.001$}
        \label{fig:tb2C}
    \end{subfigure}%
    \hfill%
    \begin{subfigure}[b]{0.075\textwidth}
        \centering
        \includegraphics[width=\textwidth, alt={Calculated Jacobi sets with Collapse Algorithm Variant A in the tensile bar D.}]{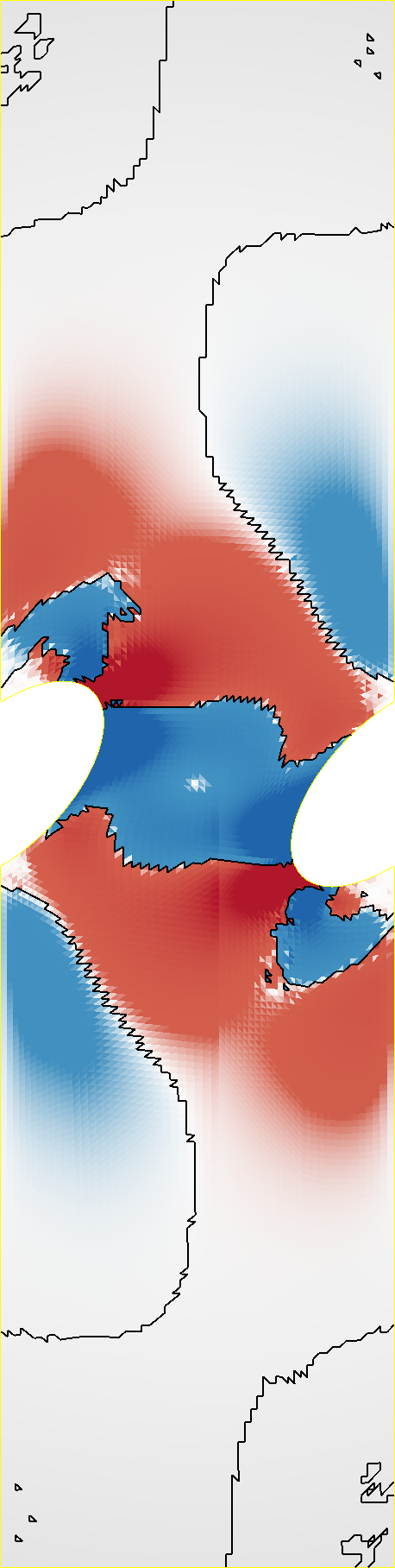}
        \caption{$t = 0.01$}
        \label{fig:tb3C}
    \end{subfigure}%
    \hfill%
    \begin{subfigure}[b]{0.10\textwidth}
        \centering
        \includegraphics[width=\textwidth, alt={Calculated Jacobi sets with Collapse Algorithm Variant A in the tensile bar E.}]{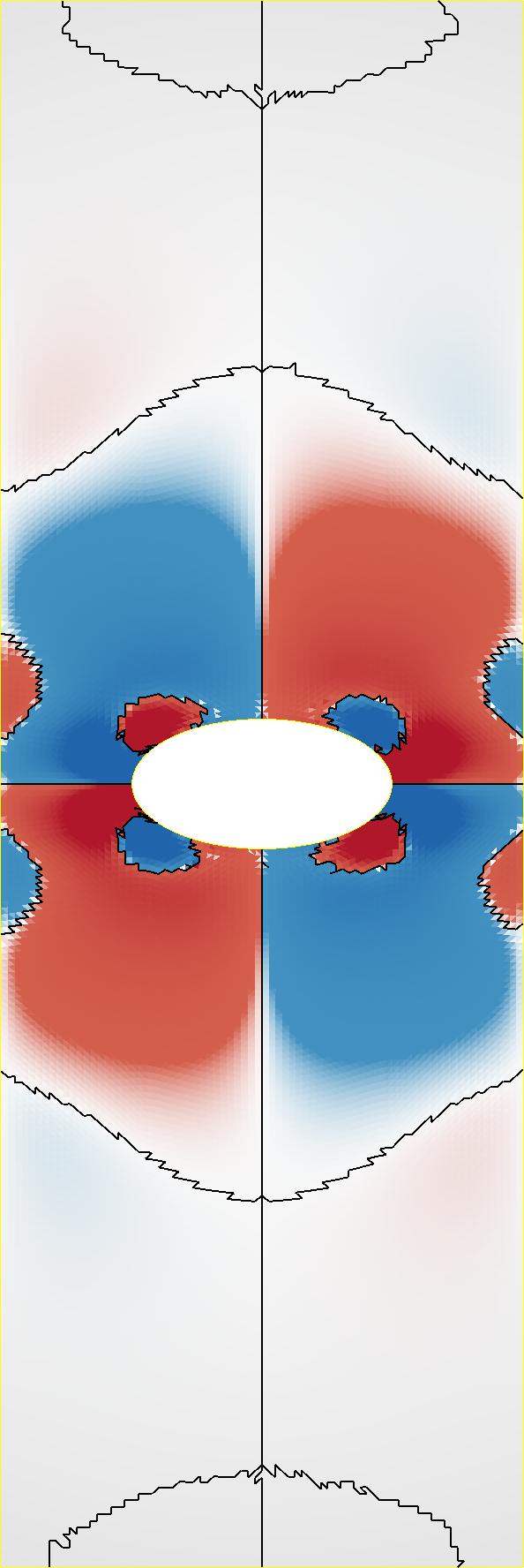}
        \caption{$t = 0.001$}
        \label{fig:tb4C}
    \end{subfigure}%
    \hfill%
    \begin{subfigure}[b]{0.075\textwidth}
        \centering
        \includegraphics[width=\textwidth, alt={Calculated Jacobi sets with Collapse Algorithm Variant A in the tensile bar F.}]{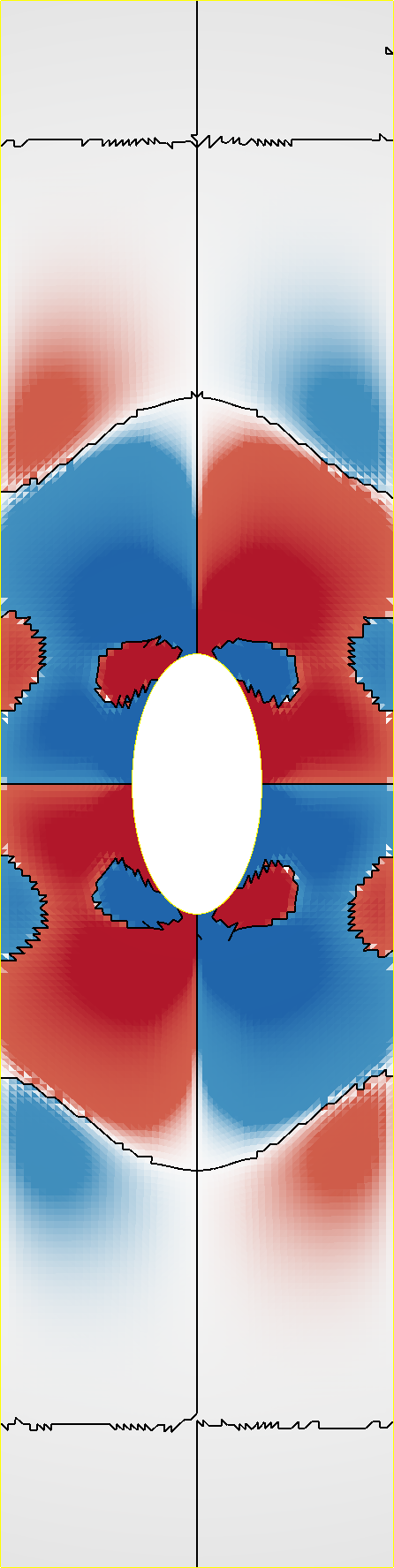}
        \caption{$t = 0.001$}
        \label{fig:tb5C}
    \end{subfigure}%
    \hfill%
    \begin{subfigure}[b]{0.10\textwidth}
        \centering
        \includegraphics[width=\textwidth, alt={Calculated Jacobi sets with Collapse Algorithm Variant A in the tensile bar G.}]{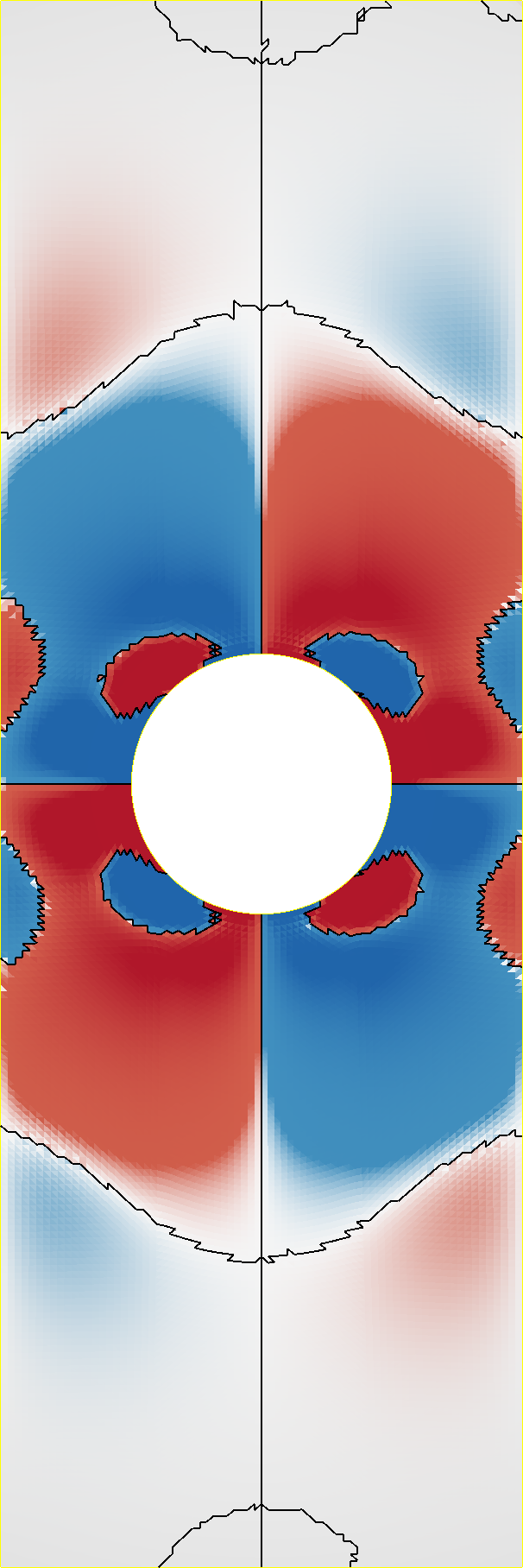}
        \caption{$t = 0.0001$}
        \label{fig:tb6C}
    \end{subfigure}%
    \hfill%
    \begin{subfigure}[b]{0.075\textwidth}
        \centering
        \includegraphics[width=\textwidth, alt={Calculated Jacobi sets with Collapse Algorithm Variant A in the tensile bar H.}]{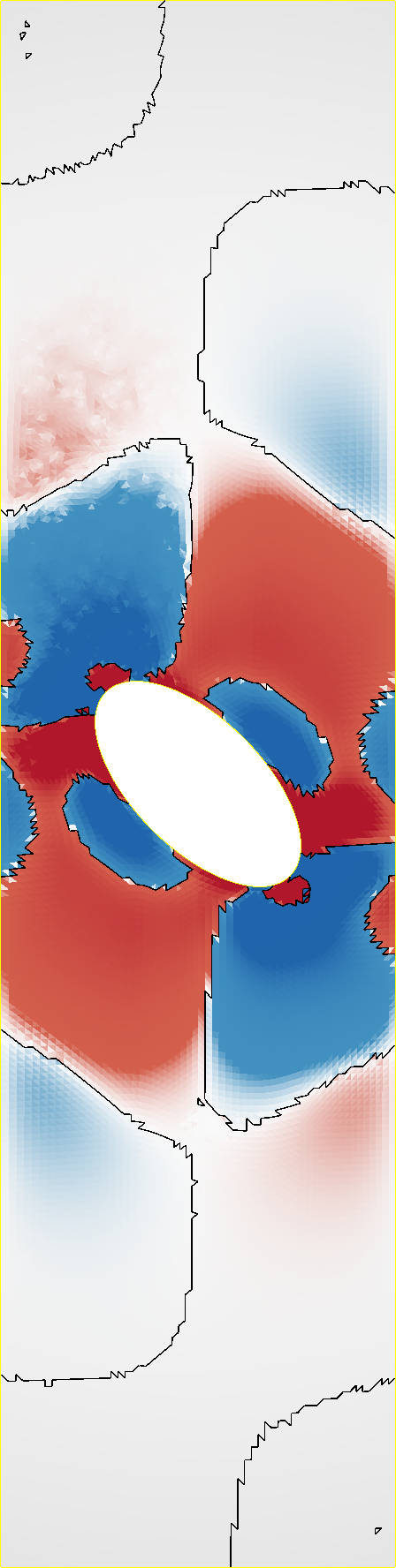}
        \caption{$t = 0.001$}
        \label{fig:tb7C}
    \end{subfigure}%
    \hspace*{\fill}
    \caption{Comparison of the calculated Jacobi sets in the Tensile Bar datasets for the original data in the upper figures and the Collapse Algorithm Variant A with the corresponding $t$ values shown in the lower figures.
}
    \label{fig:tb}
\end{figure*}

In \cref{fig:tb} we visually compare the extracted Jacobi sets in the original data as seen in the upper \cref{fig:tb0W}--h with the datasets after applying the CA variant A as seen in the \cref{fig:tb0C}--p.
The range area is displayed using color coding, with the saturation indicating how high the stresses in the dataset are.
The first thing to recognize is that the Jacobi sets in all datasets are visually simplified after applying the CA variant A and parts with small Jacobi set components could be greatly reduced.
Differences can be seen in the datasets with notches and the datasets with holes.
For example, \cref{fig:tb0W}, i shows that the Jacobi sets are simplified at the notches and the structure becomes more recognizable for domain experts.
In \cref{fig:tb2W}, k, it can be seen that in parts where a high density of small Jacobi set components occurs, the Jacobi sets are simplified, but the symmetry of the dataset is partially broken.
For example, in \cref{fig:tb4W}, m, it can be seen that the Jacobi sets are simplified in the part around the hole.
In \cref{fig:tb5W}, n it can also be seen that the Jacobi sets can be simplified at the upper and lower end of the tie rod where low stress values also appear in the dataset, which can also be seen from the white color.
In the \cref{tab:all} it can be seen that CA variant A can greatly reduce all two measures. 
The results of the loop subdivision can also be seen here.
It is noticeable that the length of all Jacobi sets increases significantly compared to the original data.
For the smoothing filters, we have no results for these datasets, as the used filters only work on structured grids.
Overall, you can see in the figures that the noise in the datasets can be greatly reduced.
The simplification from the Jacobi set components for the tensile bar D from $519$ to $40$ using the CA variant A takes $0.085$\,s.
All the other tensile bars need less time.

\subsection{Hurricane Isabel}
Another real-world dataset from the field of meteorology is \textit{Hurricane Isabel}\cite{hurricane2004isabel,klotzl2022local}
, which was published as a freely available dataset as part of the 2004 SciVis contest. 
This modeled hurricane is based on Hurricane Isabel which caused severe destruction in September 2003.
It is a 3D dataset with a resolution of $500 \times 500 \times 100$ and $48$ time steps and contains $13$ scalar variables, such as velocity, at each data point.
We use a layer from the dataset at height $50$ at time step $30$ and triangulate the grid.

\if\imageVersion2
\begin{figure*}[tb]
    \centering 
    \begin{subfigure}[b]{0.1533\textwidth}
    \begin{subfigure}[b]{\textwidth}
        \centering
        \includegraphics[width=\textwidth, alt={Calculated Jacobi sets for cutout 1 in the hurricane center of the original data of Hurricane Isabel.}]{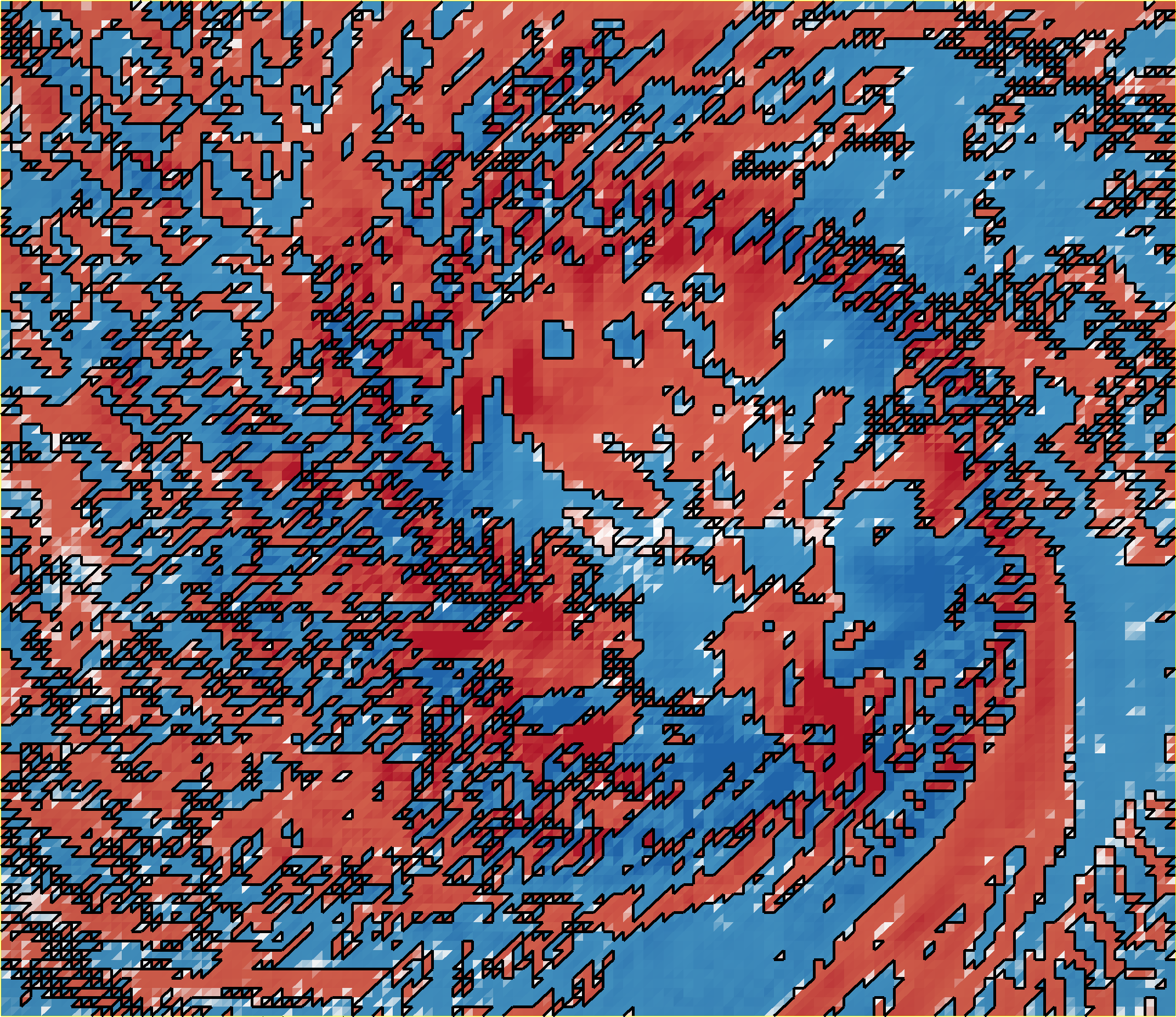}
        \caption{original Cutout 1}
        \label{fig:hiWZoom1}
    \end{subfigure}%
    \hfill%
    \begin{subfigure}[b]{\textwidth}
        \centering
        \includegraphics[width=\textwidth, alt={Calculated Jacobi sets for cutout  2 in the hurricane center of the original data of Hurricane Isabel.}]{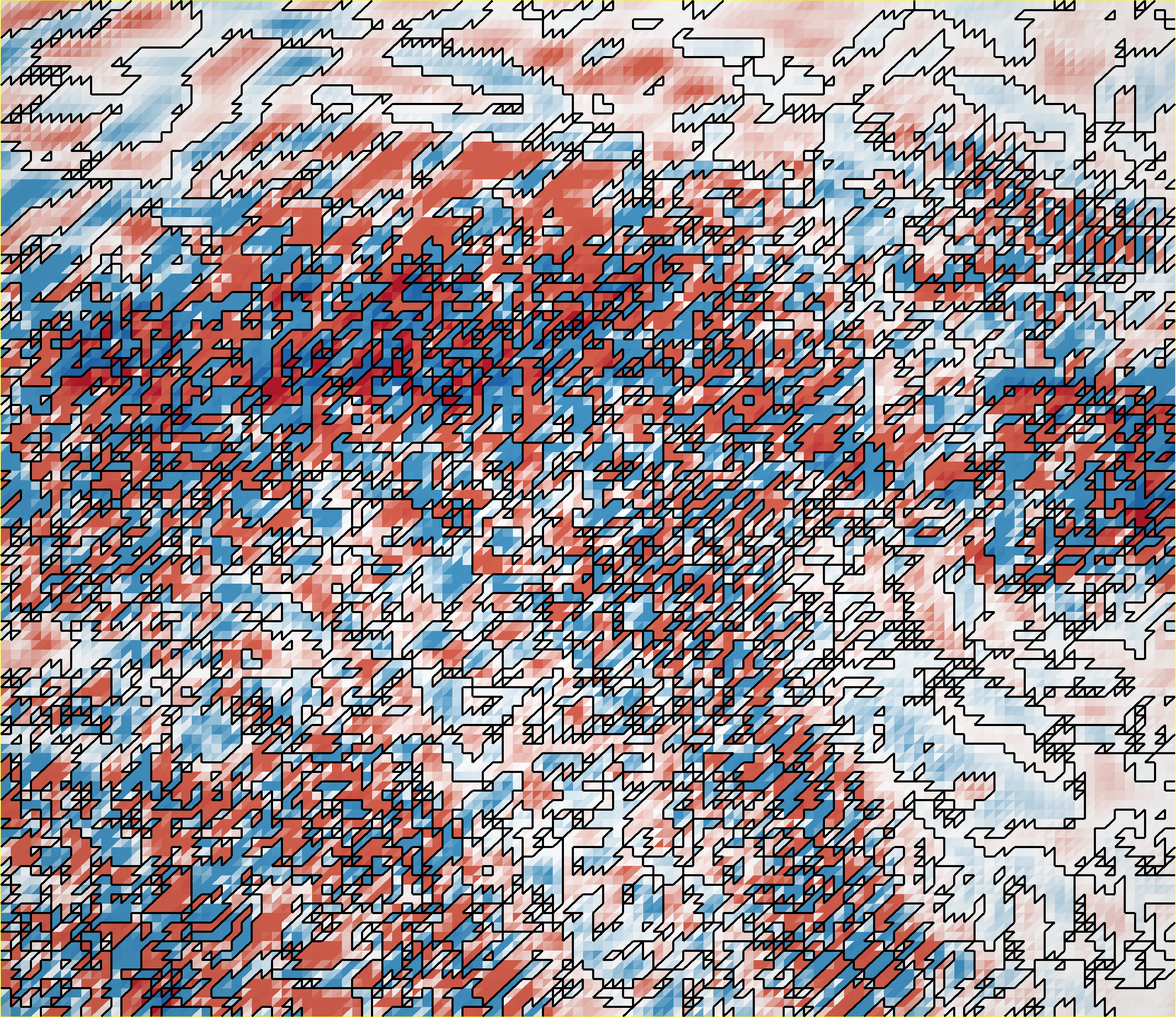}
        \caption{original Cutout 2}
        \label{fig:hiWZoom2}
    \end{subfigure}%
    \end{subfigure}%
    \hspace{0.4em}
    \begin{subfigure}[b]{0.31\textwidth}
        \centering
        \includegraphics[width=\textwidth, alt={Calculated Jacobi sets for the original data of Hurricane Isabel.}]{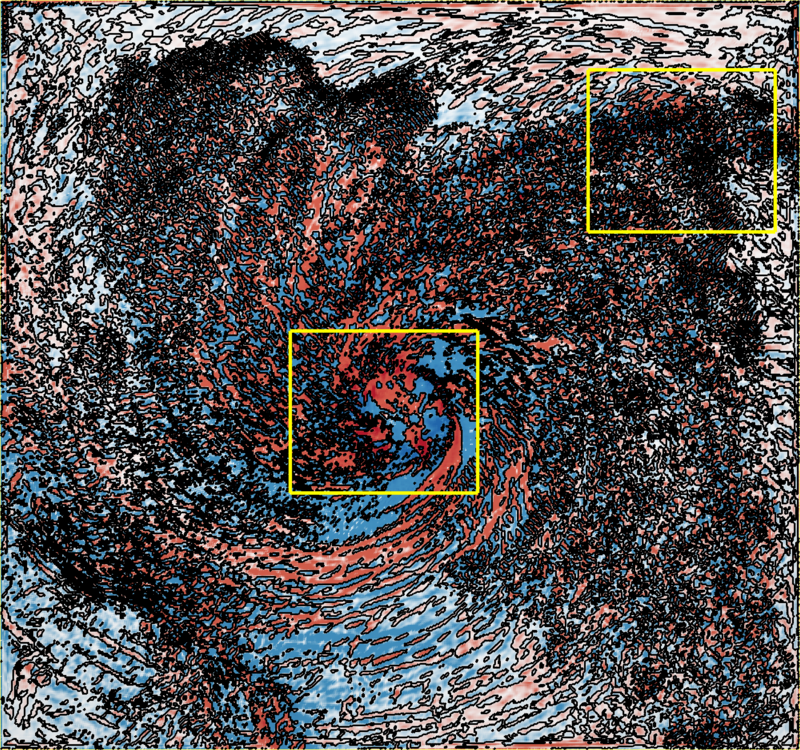}
        \caption{original}
        \label{fig:hiW}
    \end{subfigure}%
    \hfill%
    \begin{subfigure}[b]{0.31\textwidth}
    \centering
        \includegraphics[width=\textwidth, alt={Calculated Jacobi sets for Hurricane Isabel using the collapse algorithm variant A.}]{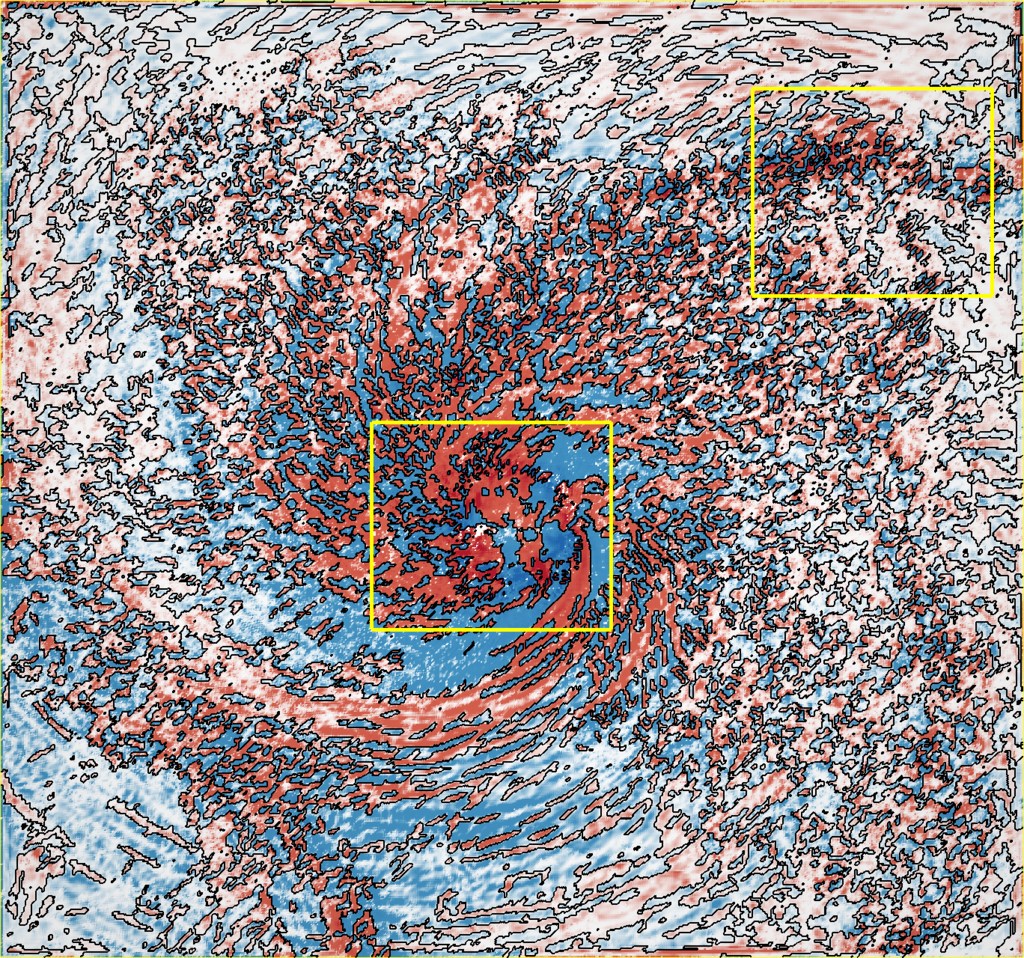}
        \caption{CA Variant A}
        \label{fig:hiC}
    \end{subfigure}%
    \hspace{0.4em}
    \begin{subfigure}[b]{0.1533\textwidth}
    \begin{subfigure}[b]{\textwidth}
        \centering
        \includegraphics[width=\textwidth, alt={Calculated Jacobi sets for Hurricane Isabel using the collapse algorithm variant A for the cutout 1 in the center of the hurricane.}]{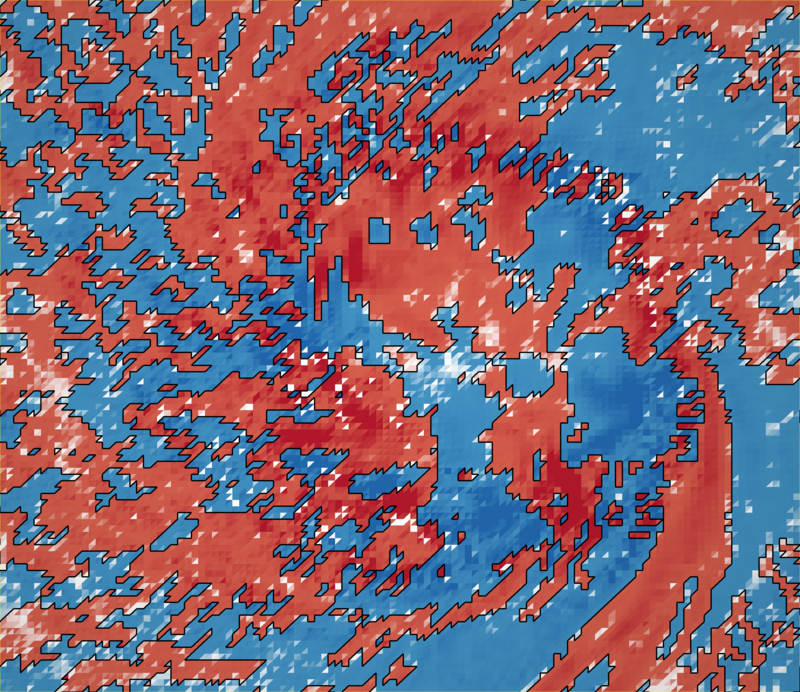}
        \caption{CA Variant A Cutout 1}
        \label{fig:hiCZoom1}
    \end{subfigure}%
    \hfill%
    \begin{subfigure}[b]{\textwidth}
        \centering
        \includegraphics[width=\textwidth, alt={Calculated Jacobi sets for Hurricane Isabel using the collapse algorithm variant A for the cutout 2 in the center of the hurricane.}]{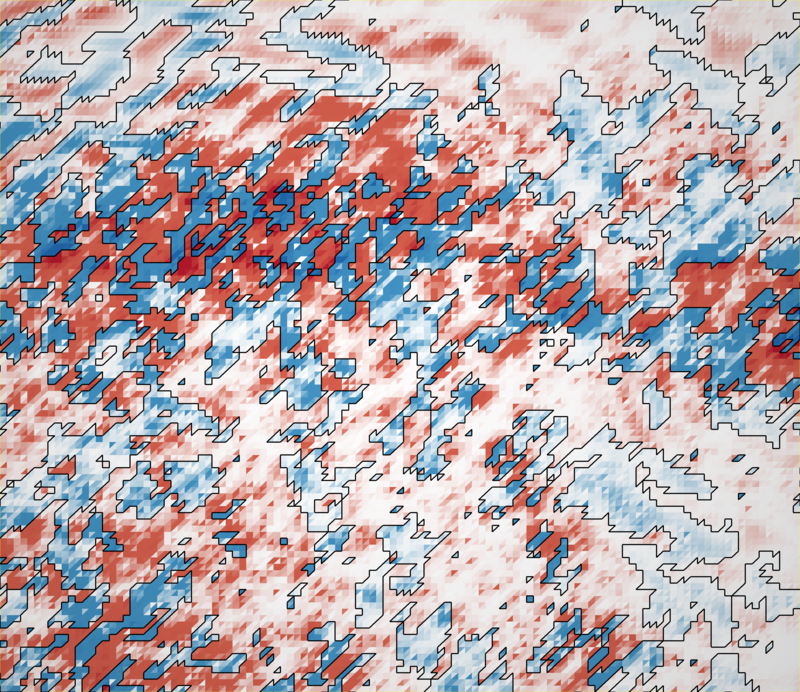}
        \caption{CA Variant A Cutout 2}
        \label{fig:hiCZoom2}
    \end{subfigure}%
    \end{subfigure}%
    \\
    \begin{subfigure}[b]{0.1533\textwidth}
    \begin{subfigure}[b]{\textwidth}
        \centering
        \includegraphics[width=\textwidth, alt={Calculated Jacobi sets for Hurricane Isabel using the Loop subdivsion algorthim for the cutout 1 in the center of the hurricane.}]{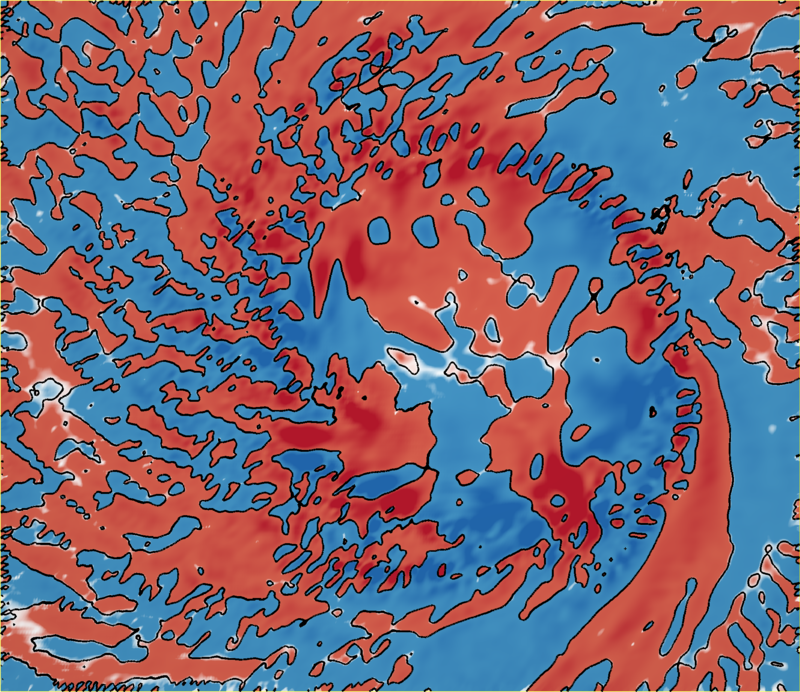}
        \caption{Loop Subdivision Cutout 1}
        \label{fig:hiL4Zoom1}
    \end{subfigure}%
    \hfill%
    \begin{subfigure}[b]{\textwidth}
        \centering
        \includegraphics[width=\textwidth, alt={Calculated Jacobi sets for Hurricane Isabel using the Loop subdivsion algorthim for the cutout 2 in the center of the hurricane.}]{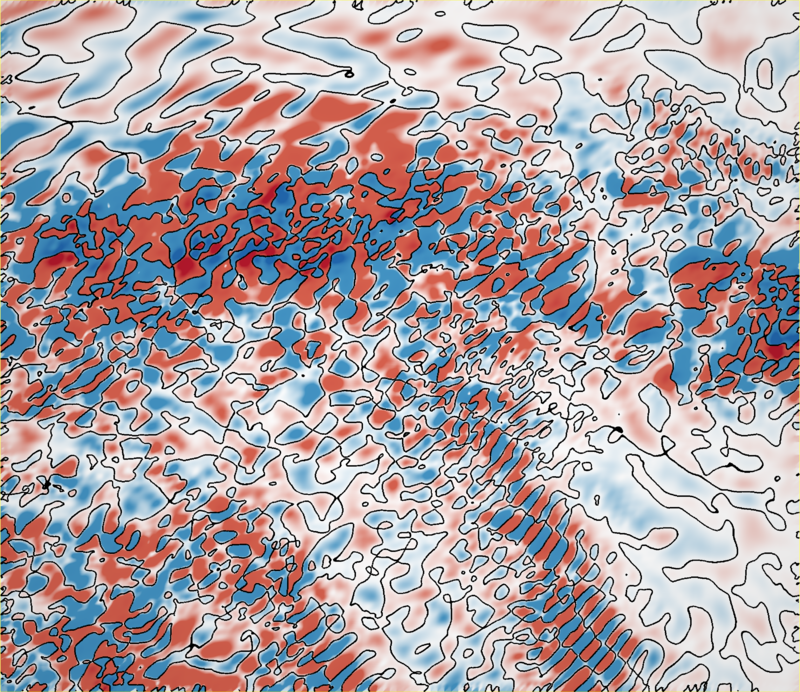}
        \caption{Loop Subdivision Cutout 2}
        \label{fig:hiL4Zoom2}
    \end{subfigure}%
    \end{subfigure}%
    \hspace{0.4em}
    \begin{subfigure}[b]{0.31\textwidth}
        \centering
        \includegraphics[width=\textwidth, alt={Calculated Jacobi sets for Hurricane Isabel using the Loop subdivision algorithm.}]{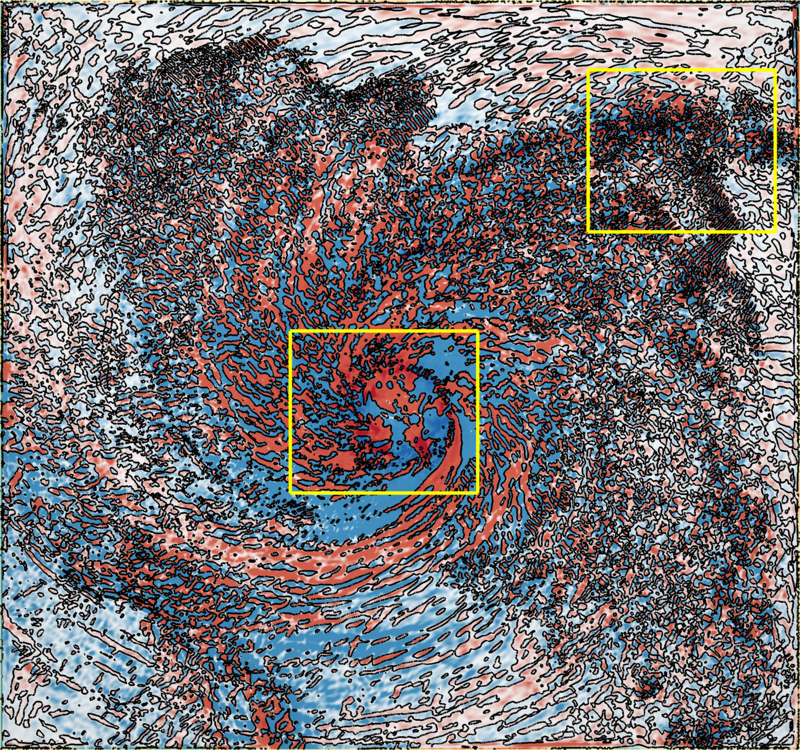}
        \caption{Loop Subdivision}
        \label{fig:hiL4}
    \end{subfigure}%
    \hfill%
    \begin{subfigure}[b]{0.31\textwidth}
    \centering
        \includegraphics[width=\textwidth, alt={Calculated Jacobi sets for Hurricane Isabel using the Gaussian filter.}]{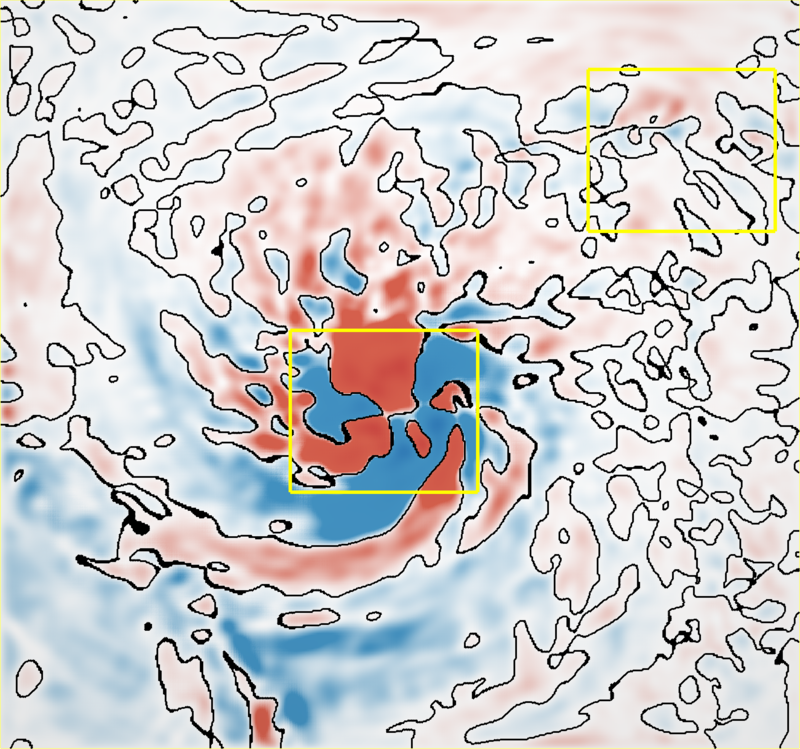}
        \caption{Gaussian filter}
        \label{fig:hiG1000}
    \end{subfigure}%
    \hspace{0.4em}
    \begin{subfigure}[b]{0.1533\textwidth}
    \begin{subfigure}[b]{\textwidth}
        \centering
        \includegraphics[width=\textwidth, alt={Calculated Jacobi sets for Hurricane Isabel using the Gaussian filter for the cutout 1 in the center of the hurricane.}]{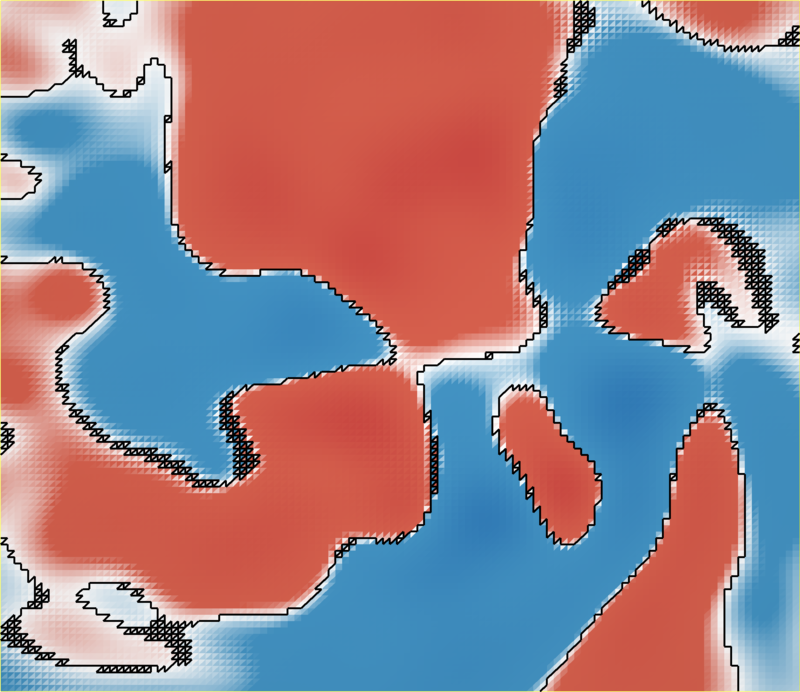}
        \caption{Gaussian filter Cutout 1}
        \label{fig:hiGZoom1}
    \end{subfigure}%
    \hfill%
    \begin{subfigure}[b]{\textwidth}
        \centering
        \includegraphics[width=\textwidth, alt={Calculated Jacobi sets for Hurricane Isabel using the Gaussian filter for the cutout 2 in the center of the hurricane.}]{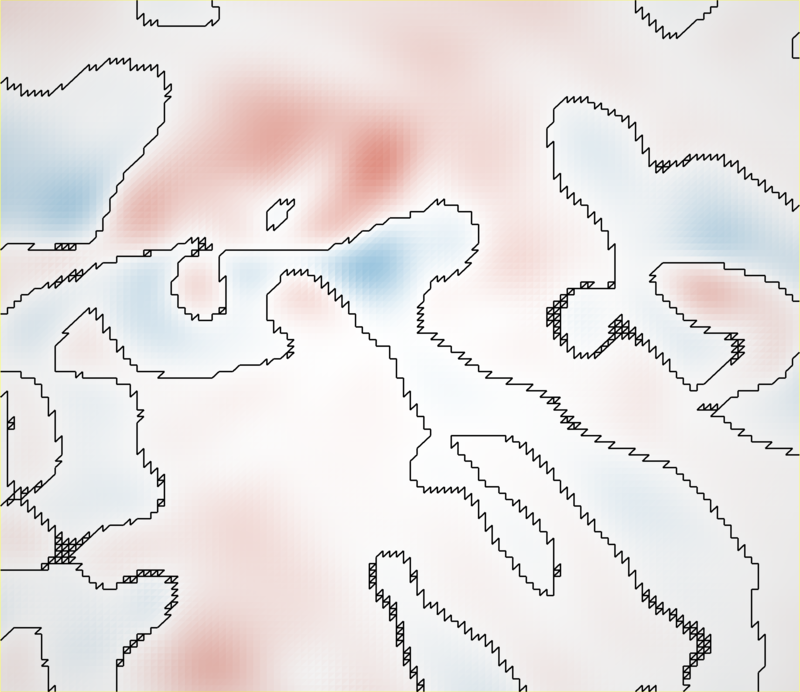}
        \caption{Gaussian filter Cutout 2}
        \label{fig:hiGZoom2}
    \end{subfigure}%
    \end{subfigure}%
    \caption{\captionHI}
    \label{fig:hi}
\end{figure*}
\fi

To do this, we look at the scalar fields for pressure and temperature for the corresponding visual analysis in \cref{fig:hi}.
This figure shows the algorithms to be compared, and the original data.
An overall view of the dataset can be seen in \cref{fig:hiW}, \ref{fig:hiC}, \ref{fig:hiL4}, and \cref{fig:hiG1000}.
Visually, it can be seen that all $3$ algorithms simplify the Jacobi sets compared to the original data.
In \cref{fig:hiWZoom1}, \ref{fig:hiCZoom1}, \ref{fig:hiL4Zoom1} and \cref{fig:hiGZoom1} a cutout of the center of the hurricane is shown.
It is easy to see that CA variant A and the loop subdivision simplify the Jacobi sets but still retain the structures.
In contrast, the Gaussian filter simplifies the center so much that it looks completely different compared to the original data.
A visual comparison in \cref{fig:hiG1000} shows that the Gaussian filter greatly simplifies many large structures. 
A second cutout of the edge of the hurricane can be seen in \cref{fig:hiWZoom2}, \ref{fig:hiCZoom2}, \ref{fig:hiL4Zoom2}, and \cref{fig:hiGZoom2}.
Here it can be seen in the original data that many small Jacobi set components occur in the part where the difference between the range areas is small.
Visually, these small Jacobi set components have decreased in the loop subdivision as well as in the CA variant A.
This also shows that the Gaussian filter greases the data too much.
\cref{tab:all} clearly shows that the CA variant A reduces the components of the Jacobi set the most and that the loop subdivision leads to more Jacobi set components despite an optical simplification. 
For the Gaussian filter, it can be seen that it reduces the components of the Jacobi sets and even reduces the length of the Jacobi sets the most, but this is at the expense of important structures.
Simplifying the Jacobi set components from $43'344$ to $2'657$ using the CA variant A takes $5.5$\,s.


\section{Discussion} 

In this study, we compared the collapse algorithm (CA) variants, smoothing filters, and a loop subdivision with the original data using different datasets to investigate the effect on the Jacobi set simplification. 
Besides, we visually examined the influence of the different neighborhood graphs on the result and compared the best variant with the smoothing filters and the loop subdivision.

Our investigations have shown that the structure of the neighborhood graph is of great importance, as the results differed from one another. 
The idea of converting the neighborhood graph into a neighborhood tree had certain advantages, as this made it possible to assign an orientation direction in particularly noisy parts. 
However, the decision as to whether the negative (variant B) or the positive orientation (variant C) was preferred led to very different results. 
Therefore, the choice of variant varies depending on the case. 
The assignment of the cells according to their local neighborhood (variant D) could already solve this problem better but led to the fact that many small Jacobi set components were still not simplified because they were assigned according to their neighborhood. 
For this reason, the neighborhood graph was used in CA variant A without further adjustments. 
This has the advantage that the structure is simpler and in the examples considered, the results were not worse, but even better. 
However, there were small noisy parts in \cref{fig:teaser} at the bottom left where not all the noise could be removed.
Compared to the original data, the results of the CA variant A are very good. 
The neighborhood graphs can be greatly simplified, which is particularly beneficial for domain experts, as the analysis becomes more difficult as the data gets larger and larger. 
This can also be seen when looking at all datasets together, whereby CA variant A was able to reduce the number of Jacobi set components by more than an order of magnitude on average compared to the original data. 
The length of the Jacobi sets and the number of Jacobi set components were reduced by more than half on average with CA variant A. 
The visual results also show that the dataset provides a significantly better overview.
The comparison of the CA variant A with the loop subdivision showed that the CA does not affect the essential structures in the dataset and can simplify the datasets at the same time. 
However, there were still deficits, particularly in the case of data with high symmetry and very noisy parts. 
Here, the CA should be further optimized to deliver even better results. 
Although the loop subdivision can visually provide very good results, the actual problem of noisy small Jacobi set components is not solved, but the triangles where this occurs become smaller, as already shown in the work of Klötzel et al. \cite{klotzl2022local} can be seen.
The comparison with the Gaussian filter, showed that although it removes common noise in the green regions of interest very well, the essential structures are also changed and thus the content is lost. 
This can also be seen in \cref{fig:hiGZoom1} where the Gaussian filter can simplify very noisy parts, leaving many small Jacobi set components.
Further comparative tests with different smoothing filters can be found in the appendix.

Overall, it can be seen that the CA variant A can greatly simplify the Jacobi sets without removing essential structures and achieves better results compared to the other algorithms.
Since the algorithm collapses cells, this can lead to the fact that a component cannot be completely removed if the algorithm runs into an oscillation, and at this point, the neighborhood graph cannot be simplified further.
Examining the runtime in all datasets, it can be seen that for CA variant A it depends on the number of Jacobi set components to be reduced and requires an average of $13.80$\.ms to reduce $100$ Jacobi set components.


\section{Conclusion \& Future Work} 
In this study, an algorithm for simplifying Jacobi sets for bivariate 2D scalar fields in unstructured triangulated grids was presented by adjusting the underlying data. 
This algorithm is based on the collapsing of cells which changes the underlying data, whereby the collapsing cells are identified using a neighborhood graph. 
This resulted in four algorithm variants.
How well the algorithm variants, smoothing filters, and loop subdivision can simplify the Jacobi sets was then studied with the original data using three datasets from the literature.
The best of the four variants was determined, which is the collapse algorithm variant A, especially with regard to the number of components of the Jacobi set, which was an order of magnitude lower than in the original data.
The Gaussian filter, on the other hand, was able to reduce the length of the Jacobi sets the most, but changed the dataset too much. 
However, the collapse algorithm was also able to reduce the length of the Jacobi sets by half.
The results of the collapse algorithm are promising, although some adjustments are still needed for symmetric data such as the tension rods to better preserve symmetry.
Besides, an adapted Jacobi set visualization was presented to account for degenerate cells and assign them to a Jacobi set component according to their neighborhood.


As indicated at various points in the paper, there is still a lot of potential for future work: 
Currently, a threshold is used to select Jacobi set components to be collapsed, here an automatic selection that works over the neighborhood could be a possible extension.
Furthermore, the presented collapse algorithm has the potential for accelerating the extraction of fiber surfaces or lines, especially if the extraction is done using Jacobi sets, as in the work of Sharma et al.~\cite{sharma2022jacobi}. 
This further development would not only increase the extraction speed but also simplify the visual analysis of the range.
Symmetrical datasets show optimization possibilities for the algorithm at various points, as the assignment of collapsed cells cannot always be unambiguous. 
One approach would be to adapt the algorithm so that cells do not collapse directly, but collapse gradually. 
This could achieve a smooth transition between the cells of the regions to be collapsed.
An obvious next step is to adapt the algorithm to work with 3D bivariate datasets, as the domain still consists of triangles.
However, this requires further investigation and evaluation of different datasets to investigate the influence on the domain.
Especially the selection of the Jacobi set components to be collapsed can be difficult. 
Furthermore, the focus is on extending the algorithm for 3D case.
Here, it would first have to be investigated how the collapsing of tetrahedra can be realized and what effects this has on the dataset. 
This would introduce a new dimension of complexity and require careful investigation and testing with different datasets.

\acknowledgments{%
    This work was funded by the Deutsche Forschungsgemeinschaft (DFG, German Research Foundation) - SCHE 663/17-1.
    The authors would like to thank Markus Stommel from Leibniz Institute of Polymer Research Dresden, Germany for providing the tensile bar datasets.
    The authors will like to thank Bill Kuo, Wei Wang, Cindy Bruyere, Tim Scheitlin, and Don Middleton of the U.S. National Center for Atmospheric Research (NCAR), and the U.S. National Science Foundation (NSF) for providing the Weather Research and Forecasting (WRF) Model simulation data of Hurricane Isabel.
    The authors acknowledge the financial support by the Federal Ministry of Education and Research of Germany and by Sächsische Staatsministerium für Wissenschaft, Kultur und Tourismus in the programme Center of Excellence for AI-research ``Center for Scalable Data Analytics and Artificial Intelligence Dresden/Leipzig'', project identification number: SCADS24B.
}

\bibliographystyle{abbrv-doi}

\bibliography{template}

\clearpage


\section*{Supplementary material}

# Topological Simplifcation of Jacobi Sets for Piecewise-Linear Bivariate 2D Scalar Fields with Adjustment of the Underlying Data (Supplement)

Felix Raith [ID] *
Leipzig University

Gerik Scheuermann [ID] †
Leipzig University & ScaDS.AI

Christian Heine [ID] ‡
Leipzig University

## A  EXTENDED RESULTS

In this section, we present detailed results of our study. We show additional comparison for the Gaussian filter with a $\sigma \in \{10, 50, 100, 500\}$ and give a detailed overview of the comparative tests in the extended Tab. 2. We consider another metric in the extended table, the number of Jacobi set segments. This metric can be relevant, especially when Jacobi sets are used in acceleration structures to enable a faster selection of the areas to be considered.

In the Tab. 1 a runtime analysis is given for the collapse algorithm variant A. The demonstrations were run on a MacBook Pro (14-inch, 2021) with a Apple M1 Max, and 64 GB Ram. Implemented is the algorithm inside our framework with C++. The time measurements show that the algorithm is already interactive for most datasets but that there is still potential for acceleration for very complex data sets. However, this is not so important as the algorithm counts more than one preprocessing step. Furthermore, it turns out that the time depends on the components of the Jacobi set to be reduced.

Table 1: Time analysis of the Collapse Algorithm Variant A for all datasets. All times are given in ms.

| Dataset | Overall time | Reduced jacobi set components | Time to reduce 100 jacobi set components |
|---|---|---|---|
| Cylinder Flow | 211.07 | 662 | 31.88 |
| Tensile Bar A | 58.74 | 798 | 7.36 |
| Tensile Bar B | 60.81 | 750 | 8.11 |
| Tensile Bar C | 54.38 | 865 | 6.29 |
| Tensile Bar D | 85.82 | 479 | 17.92 |
| Tensile Bar E | 52.47 | 452 | 11.61 |
| Tensile Bar F | 36.44 | 305 | 11.95 |
| Tensile Bar G | 38.17 | 290 | 13.16 |
| Tensile Bar H | 48.42 | 294 | 16.47 |
| Hurricane Isabel | 5'461.22 | 41181 | 13.26 |

Overall, the application of the Gaussian filter with different $\sigma$ values shows that important structures, as can be seen in the center of the hurricane (see Fig. 6a), are reduced.

### A.1  Cylinder Flow (Synthetic)

The section contains corresponding figures for the different Gaussian filters. The visual analysis shows that the Gaussian filters with

*e-mail: raith@informatik.uni-leipzig.de
†e-mail: scheuer@informatik.uni-leipzig.de
‡e-mail: heine@informatik.uni-leipzig.de

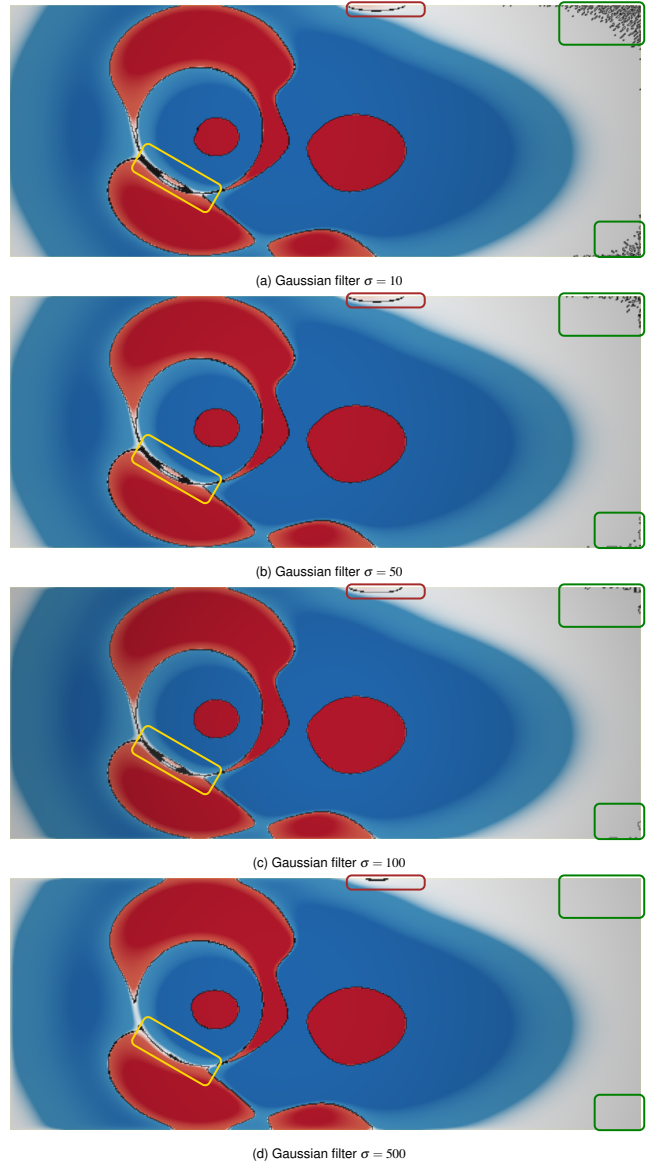

Figure 1: Comparison of the calculated Jacobi sets in the Cylinder Flow dataset for the Gaussian filter $\sigma = 10$ (a), $\sigma = 50$ (b), $\sigma = 100$ (c), and $\sigma = 500$ (d).

$\sigma \in \{10, 50, 100\}$ already have problems reducing all Jacobi set components from the first green region of interest (ROI). In the other two ROIs, the reduction is also not sufficient to remove the small Jacobi set components. For the Gaussian filter with $\sigma = 500$, all Jacobi set components can be removed in the first green ROI. In the second brown ROI, the large Jacobi set component is already greatly

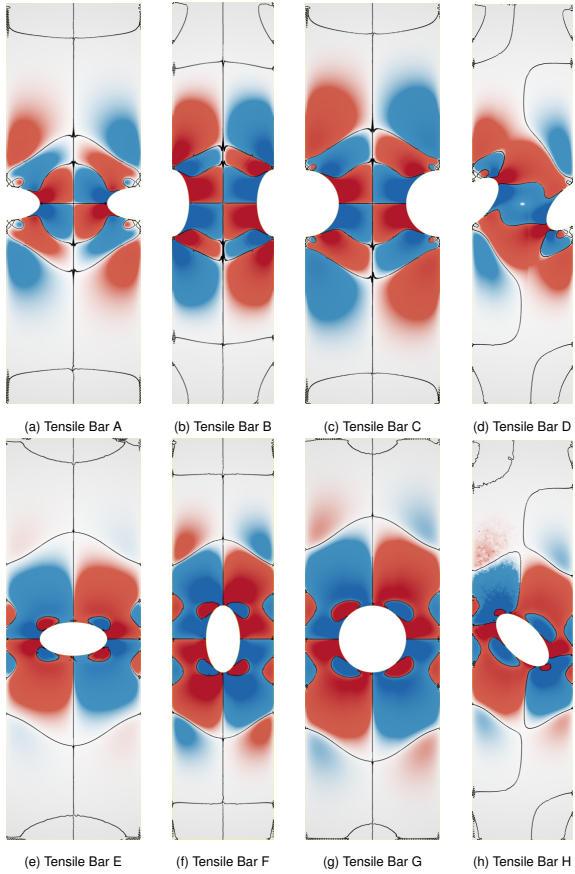

Figure 2: Calculated Jacobi sets in the Tensile Bar datasets for the Loop subdivision Algorithm with 4 subdivision steps.

reduced in size, and the small ones cannot be removed. A similar picture emerges in the third yellow ROI in which the larger Jacobi set components have already been removed, and only many small Jacobi set components can be seen in their place.

### A.2 Tensile Bar

To fully analyze the effectiveness of the loop subdivision to simplify the Jacobi sets, we give the visual results of applying the loop subdivision to the tensile bars. Here can see that the Jacobi sets can be simplified visually for the different tensile bars. In detail, however, it can be seen that there are some problems. For example, in tensile bar G in Fig. 2g it is possible to find regions in which more new Jacobi set components have formed. Such areas can also be seen in the other tensile bars, especially where Jacobi sets intersect in the original data. These visual results confirm the results from the Tab. 2, in which the length of the Jacobi sets in the tensile bars is greater than the length of the original data, in contrast to the cylinder flow and Hurricane Isabel data sets. It can also be seen in the Tab. 2 that the number of Jacobi set components increases much more than in the two other data sets before. This could lead to a misinterpretation of the visual analysis.

### A.3 Hurricane Isabel

In Fig. 3a, Fig. 4a, Fig. 5a, Fig. 6a, and Fig. 7a an overall view of the hurricane Isabelle dataset is shown after applying the binominal filter and the gaussian filter with $\sigma \in \{10, 50, 100, 500\}$. These smoothing filters also show visually that the Jacobi set components can be simplified visually. In detail, the cutout of Fig. 3b, Fig. 4b, and Fig. 5b shows that the reduction is present in the center of

the hurricane and that the structures remain largely intact when compared with the original data. In the Fig. 6b, and Fig. 7b, the filter effect is already too strong, and important structures are no longer fully recognizable. The cutout in Fig. 3c, Fig. 4c, Fig. 5c, Fig. 6c, and Fig. 7c provides a slightly different picture, whereby a real simplification and thus the recognition of structures has not been improved. This is particularly evident in the comparison with the loop subdivision and the CA variant A. The image of the visual analysis is confirmed in the values from the Tab. 2, whereby a reduction of the Jacobi set components can be recognized and thus also a reduction of the length of the Jacobi sets themselves. It turns out that these smoothing filters are also not better suited for simplifying the Jacobi set components.

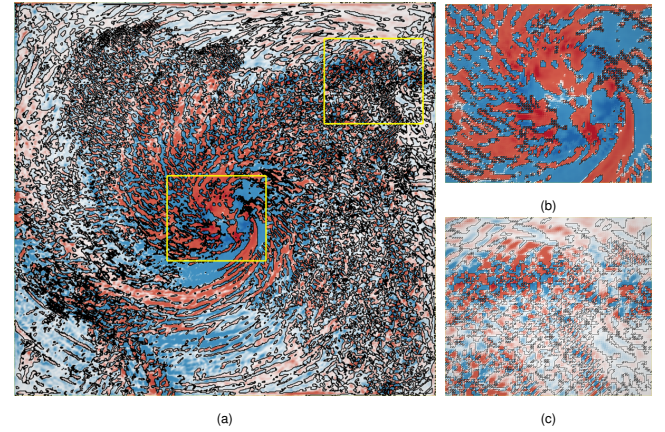

Figure 3: The calculated Jacobi sets in the Hurricane Isabel dataset for the Binomial filter (a) and additional in detail for cutout 1 (coordinates (772, 683) − (1278, 1122)) (b) and cutout 2 (coordinates (1572, 1382) − (2077, 1822)) (c).

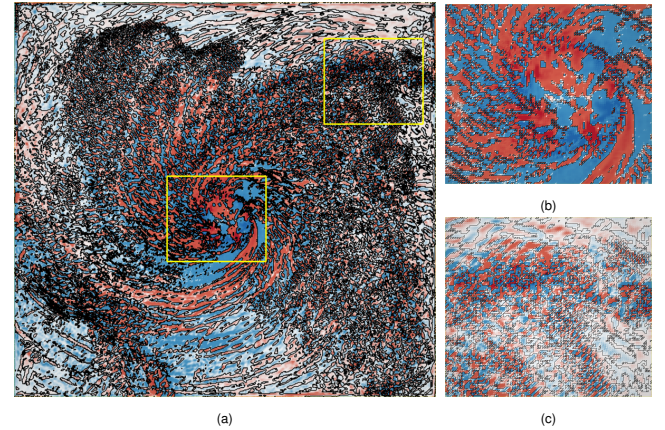

Figure 4: The calculated Jacobi sets in the Hurricane Isabel dataset for the Gaussian filter with $\sigma = 10$ (a) and additional in detail for cutout 1 (coordinates (772, 683) − (1278, 1122)) (b) and cutout 2 (coordinates (1572, 1382) − (2077, 1822)) (c).

Table 2: The results for the length of the Jacobi sets, the number of separated Jacobi set segments, and the number of Jacobi set components in detail can be seen for all datasets and methods tested, with the bolded values being the best.

| Dataset | Method | Cells | Length of all Jacobi sets | # of Jacobi set segments | # of separated Jacobi sets components |
|---|---|---|---|---|---|
| Cylinder Flow (jungtelziemniak2d) | Original Data | 178'702 | 92.4614 | 3839 | 679 |
| | Binominal filter | 178'702 | 64.9239 | 2746 | 448 |
| | Gauss filter with $\sigma = 10$ | 178'702 | 79.4928 | 3'329 | 564 |
| | Gauss filter with $\sigma = 50$ | 178'702 | 61.1041 | 2'393 | 405 |
| | Gauss filter with $\sigma = 100$ | 178'702 | 56.3566 | 2'393 | 353 |
| | Gauss filter with $\sigma = 500$ | 178'702 | 43.8264 | 1'872 | 143 |
| | Gauss filter with $\sigma = 1000$ | 178'702 | **40.0536** | **1'724** | 101 |
| | Loop subdivision Algorithm | 45'747'712 | 82.9961 | 55'192 | 6'311 |
| | Collapse Algorithm Variant A | 178'702 | 44.056 | 1'879 | **17** |
| | Collapse Algorithm Variant B | 178'702 | 48.8755 | 2'073 | 41 |
| | Collapse Algorithm Variant C | 178'702 | 50.2173 | 2'126 | 42 |
| | Collapse Algorithm Variant D | 178'702 | 49.8907 | 2'121 | 23 |
| Tensile Bar A | Original Data | 30'960 | 1'382.95 | 2'349 | 817 |
| | Loop subdivision Algorithm | 79'25'760 | 1'845.95 | 50'018 | 17'858 |
| | Collapse Algorithm Variant A | 30'960 | **615.601** | **1'067** | **19** |
| Tensile Bar B | Original Data | 22'360 | 1'476.09 | 2'289 | 800 |
| | Loop subdivision Algorithm | 5'724'160 | 2'200.9 | 55'415 | 23'625 |
| | Collapse Algorithm Variant A | 22'360 | **767.146** | **1'182** | **50** |
| Tensile Bar C | Original Data | 29'560 | 1'503.17 | 2'439 | 883 |
| | Loop subdivision Algorithm | 7'567'360 | 2'068.83 | 53'992 | 21'529 |
| | Collapse Algorithm Variant A | 29'560 | **675.679** | **1'103** | **18** |
| Tensile Bar D | Original Data | 26'850 | 963.471 | 1'681 | 519 |
| | Loop subdivision Algorithm | 6'873'600 | 1'010.44 | 28'336 | 7'164 |
| | Collapse Algorithm Variant A | 26'850 | **465.583** | 829 | **40** |
| Tensile Bar E | Original Data | 36'336 | 1'004.65 | 1'776 | 490 |
| | Loop subdivision Algorithm | 9'302'016 | 1'238.53 | 35'312 | 9'442 |
| | Collapse Algorithm Variant A | 36'336 | **535.383** | 980 | **38** |
| Tensile Bar F | Original Data | 27'210 | 791.891 | 1'392 | 329 |
| | Loop subdivision Algorithm | 6'965'760 | 1'004.82 | 28'972 | 7'259 |
| | Collapse Algorithm Variant A | 27'210 | **478.767** | 870 | **24** |
| Tensile Bar G | Original Data | 34'012 | 815.444 | 1'426 | 326 |
| | Loop subdivision Algorithm | 8'707'072 | 1'094.49 | 30'530 | 6'827 |
| | Collapse Algorithm Variant A | 34'012 | **514.535** | 926 | **36** |
| Tensile Bar H | Original Data | 28'226 | 782.771 | 1'399 | 322 |
| | Loop subdivision Algorithm | 7'225'396 | 846.473 | 24'196 | 4'484 |
| | Collapse Algorithm Variant A | 28'226 | **447.887** | 820 | **28** |
| Hurricane Isabel | Original Data | 498'002 | 915'064 | 194'923 | 43'838 |
| | Binominal filter | 498002 | 662'059 | 141'847 | 27'344 |
| | Gauss filter with $\sigma = 10$ | 498'002 | 815'635 | 174'267 | 36'244 |
| | Gauss filter with $\sigma = 50$ | 498'002 | 572'704 | 122'935 | 22'068 |
| | Gauss filter with $\sigma = 100$ | 498'002 | 427'678 | 91'946 | 15'905 |
| | Gauss filter with $\sigma = 500$ | 498'002 | 181'582 | 39'165 | 6'979 |
| | Gauss filterwith $\sigma = 1000$ | 498'002 | **122'871** | **26'524** | 4'832 |
| | Loop subdivision Algorithm | 127'488'512 | 801'508 | 2'737'567 | 336'823 |
| | Collapse Algorithm Variant A | 498'002 | 393'343 | 84'089 | **2'657** |

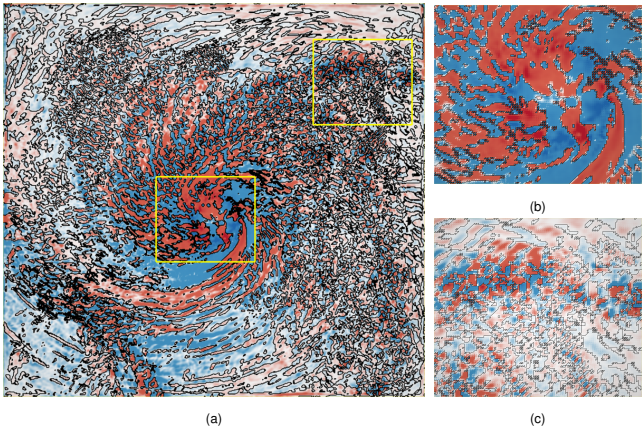

Figure 5: The calculated Jacobi sets in the Hurricane Isabel dataset for the Gaussian filter with $\sigma = 50$ (a) and additional in detail for cutout 1 (coordinates $(772, 683) - (1278, 1122)$) (b) and cutout 2 (coordinates $(1572, 1382) - (2077, 1822)$) (c).

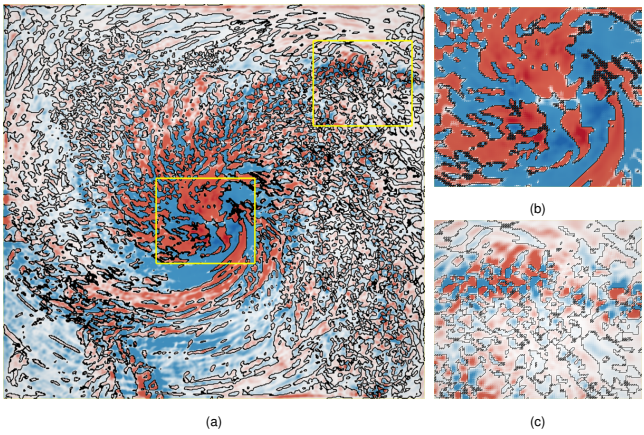

Figure 6: The calculated Jacobi sets in the Hurricane Isabel dataset for the Gaussian filter with $\sigma = 100$ (a) and additional in detail for cutout 1 (coordinates $(772, 683) - (1278, 1122)$) (b) and cutout 2 (coordinates $(1572, 1382) - (2077, 1822)$) (c).

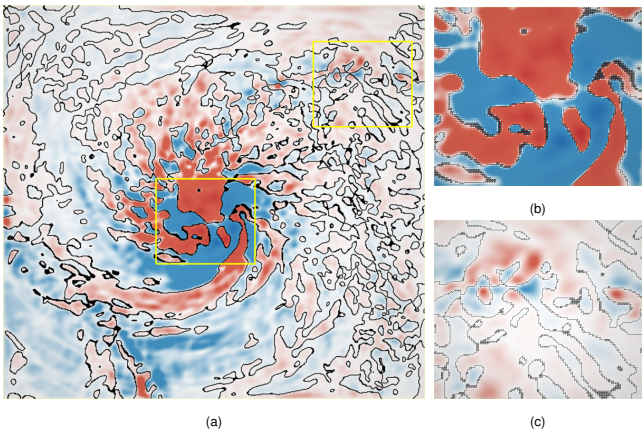

Figure 7: The calculated Jacobi sets in the Hurricane Isabel dataset for the Gaussian filter with $\sigma = 500$ (a) and additional in detail for cutout 1 (coordinates $(772, 683) - (1278, 1122)$) (b) and cutout 2 (coordinates $(1572, 1382) - (2077, 1822)$) (c).



\end{document}